\documentclass{article}

\usepackage{arxiv}
\usepackage{graphicx}
\usepackage[utf8]{inputenc} % allow utf-8 input
\usepackage[T1]{fontenc}    % use 8-bit T1 fonts
\usepackage{hyperref}       % hyperlinks
\usepackage{url}            % simple URL typesetting
\usepackage{booktabs}       % professional-quality tables
\usepackage{amsfonts}       % blackboard math symbols
\usepackage{nicefrac}       % compact symbols for 1/2, etc.
\usepackage{microtype}      % microtypography
\usepackage{lipsum}

\usepackage{authblk}
\usepackage{booktabs} % For formal tables
\usepackage{caption}
\usepackage{subfig}
\usepackage{url}
\usepackage{epstopdf}
\usepackage{array}
\usepackage{pifont}
\usepackage{multirow}
\usepackage{amssymb}
\usepackage{footnote}
\usepackage{tablefootnote}

\usepackage{color}

\newcommand{\new}[1]{\textcolor{blue}{#1}}

\usepackage{amsmath}
\usepackage{algpseudocode,algorithm,algorithmicx}

\makeatletter
\def\BState{\State\hskip-\ALG@thistlm}
\makeatother

\title{An Unsupervised Normalization Algorithm for Noisy Text: A Case Study for Information Retrieval and Stance Detection}
\date{}

\author[1]{Anurag Roy}
\author[1]{Shalmoli Ghosh}
\author[2]{Kripabandhu Ghosh}
\author[1]{Saptarshi Ghosh}

\affil[1]{Department of Computer Science and Engineering\\
Indian Institute of Technology Kharagpur \\
  Kharagpur \\
  India
}
\affil[2]{Department of Computer Science and Application \\
 Indian Institute of Science Education and Research Kolkata \\
 Mohanpur \\
India}
\affil[ ]{\textit { \{anu15roy, shalmolighosh94, kripa.ghosh\}@gmail.com, saptarshi@cse.iitkgp.ac.in}}
\newcommand{\model}{\texttt{UnsupClean}}
\date{}
\begin{document}
\date{}
\maketitle

\begin{abstract}
  A large fraction of textual data available today contains various types of `noise', such as OCR noise in digitized documents, noise due to informal writing style of users on microblogging sites, and so on. 
  To enable tasks such as search/retrieval and classification over all the available data, we need robust algorithms for text normalization, i.e., for cleaning different kinds of noise in the text. There have been several efforts towards cleaning or normalizing noisy text;
  however, many of the existing text normalization methods are supervised, and require language-dependent resources or large amounts of training data that is difficult to obtain. 
  We propose an unsupervised algorithm for text normalization that does not need any training data / human intervention. The proposed algorithm is applicable to text over different languages, and can handle both machine-generated and human-generated noise. 
  Experiments over several standard datasets show that text normalization through the proposed algorithm enables better retrieval and stance detection, as compared to that using several baseline text normalization methods. Implementation of our algorithm can be found at \url{ https://github.com/ranarag/UnsupClean}.
\end{abstract}
% \unmarkedfntext{\textcolor{blue}{\bf This work will be appearing in ACM Journal of Data and Information Quality.}}  
{\renewcommand*\@{}
\footnotetext{\textcolor{blue}{\bf This work will be appearing in ACM Journal of Data and Information Quality.}}
\makeatother}

% keywords can be removed
% \keywords{First keyword \and Second keyword \and More}
% \input{intro}
% \input{related}
% \input{approach}
% \input{baselines}
% \input{expt-retrieval}
% \input{expt-stance}
% \input{conclu}

\section{Introduction and Motivation}
% \vspace{-3mm}

%The presence of noise in text collections hurts supervised natural language processing (NLP) and Information Retrieval (IR) performance considerably.

Digitization and the advent of Web 2.0 have resulted in a staggering increase in the amount of online textual data. 
As more and more textual data is generated online, or digitized from offline collections, the presence of noise in the text is a growing concern.
Here {\it noise} refers to the incorrect forms of a valid word (of a given language). Such noise can be of two broad categories~\cite{Subramaniam-noisy-text-survey}: 
(1)~{\it machine-generated}, such as those produced by Optical Character Recognition (OCR) systems, Automatic Speech Recognition (ASR) systems, etc. while digitizing content, and 
(2)~{\it human-generated}, produced by the casual writing style of humans, typographical errors, etc. For instance, social media posts (e.g. microblogs) and SMS contain frequent use of non-standard abbreviations (e.g., `meds' for `medicines', `tmrw' for `tomorrow').

There exist plenty of important resources of textual data
%, worthy of efficient retrieval, classification 
containing both the categories of noise. 
With regard to machine-generated noise, old documents like legal documents, defense documents, collections produced from large scale book digitization projects\footnote{\url{https://archive.org/details/millionbooks}} etc. are either in scanned version or in hard-copy format. Digitization of these vital collections involve OCR-ing the scanned versions. 
%For instance, the IIT CDIP collection used in the TREC Legal Track\footnote{\url{https://trec-legal.umiacs.umd.edu/}} contains 7 million documents produced by scanning and OCRing hard-copies. 
The print-quality, font variability, etc. lead to poor performance of the OCR systems which incorporates substantial error in the resulting text documents. Moreover, these noisy documents can be present in many languages, which also contributes to the challenge, since the OCR systems of many languages may not be well-developed. The unavailability of error-free versions of these collections makes error modelling infeasible. 

For the second category of noise (user-generated), social networking sites such as Twitter and Weibo have emerged as effective resources for important real-time information in many situations~\cite{Basu2017ANW,Teevan-twitter-search, mohammad2016semeval}.
Informal writing by users in such online media generates noise in the form of spelling variations, arbitrary shortening of words, etc.~\cite{roy-cikm17}. 

%There exist variations of valid words which do not follow any grammatical rules, due to spelling variations, arbitrary shortening of words, etc.~\cite{cikmroy2017}.

The presence of various types of noise are known to adversely affect Information Retrieval (IR), Natural Language Processing (NLP) and Machine Learning (ML) applications such as search/retrieval, classification, stance detection, and so on~\cite{Vinciarelli-noisy-text-categorize}.
Hence, normalization/cleansing of such noisy text has a significant role in improving performance of downstream NLP / IR / ML tasks. 
The soul of a text normalization technique lies in mapping the different variations of a word to a single `normal' form, achieving which successfully can lead to remarkable improvement in a lot of downstream tasks. 
There have been several works that have attempted text normalization, both in  supervised and unsupervised ways (see Section~\ref{sec:related} for a survey).
Supervised algorithms require large training data (e.g., parallel corpora of noisy and clean text) which is expensive to obtain. This leads to a necessity for unsupervised methods, on which we focus in this paper. 

There already exist several unsupervised text normalization approaches. However, most of the existing works have experimented with only one type of noise. 
For instance, Ghosh et al.~\cite{ghosh-ipm} developed a method for correcting OCR errors, while others~\cite{rangarajan-sridhar-2015-unsupervised,bertaglia2016exploring} addressed variations in tweets. 
The type of errors in various text collections can be quite different -- usually in tweets the error {\it does not} occur at the beginning of a word. However, OCR errors can appear at any location in a word, which makes noisy text normalization more challenging. 
So, a robust text normalization system for generic noise should be adept at handling many random and unseen error patterns. 
But to our knowledge, no one has attempted to address different categories of noise at the same time.

In this paper, we propose a novel unsupervised, language-independent algorithm -- called \model{} -- capable of automatically identifying the noisy variants produced by any of the aforementioned categories of noise, viz. machine-generated and human-generated. 

We perform extensive experiments where we compare the proposed \model{} algorithm with four baseline text normalization algorithms -- Aspell, Enelvo~\cite{bertaglia2016exploring}, and the algorithms developed by  Sridhar et al~\cite{rangarajan-sridhar-2015-unsupervised} and Ghosh et al.~\cite{ghosh-ipm}.
To compare the performance of the various text normalization algorithms, we focus on {\it two practical downstream applications -- retrieval/search, and stance detection}. We judge the performance on a text normalization algorithm by checking the performance of state-of-the-art retrieval/stance detection models on datasets cleaned by the normalization algorithm.
Experiments are performed over several standard datasets spanning over two languages (English and Bengali), both containing OCR noise, as well as over microblogs containing human-generated noise.
The empirical evaluation demonstrates the efficacy of our proposed algorithm over the baseline normalization algorithms -- while we produce significantly better performance on OCR collections, our performance is competitive with all the baselines over the microblog collections.
To our knowledge, no prior work on text normalization/cleansing have reported such detailed comparative evaluation as done in this work -- over two downstream applications, and over six datasets spanning two languages and both machine-generated and user-generated noise.

To summarize, the strengths of the proposed text normalization algorithm \model{} are as follows:
(1)~The algorithm is completely unsupervised, and does not need any training data (parallel corpus), dictionaries, etc. Thus, the algorithm is suitable for application on resource-poor languages. 
(2)~The algorithm can handle different types of noise, including both intentional mis-spellings by human users, as well as machine-generated noise such as OCR noise, and
(3)~The algorithm is language-independent, and can be readily applied to text in different languages. 
We make the implementation of \model{} publicly available at \url{https://github.com/ranarag/UnsupClean}.

%Our proposed model relies on only one hyper-parameter ($\alpha$ ) unlike Ghosh et. al ~\cite{ghosh-ipm} which requires two hyper-parameters to be tuned making it easier to apply on noisy documents from different sources in comparison to Ghosh et. al. Our proposed model is language independent which makes it capable to be used on noisy documents containing low-resource language.

% Our model relies on only one hyper-parameter ($\alpha$ ) unlike Ghosh et. al ~\cite{ghosh-ipm} and does not rely on on an external clean corpus for list of canonical words like Enelvo et. al.~\cite{bertaglia2016exploring} and Shridhar et. al ~\cite{}. Our model is language independent and does not require any dictionary unlike Aspell~\cite{rangarajan-sridhar-2015-unsupervised}

The remainder of the paper is organized as follows. Section~\ref{sec:related} briefly surveys various types of text normalization/cleaning algorithms. 
%We describe the datasets used for experiments in Section \ref{sec:data}. 
The proposed text normalization algorithm (\model{}) is detailed  in Section~\ref{sec:approach}. 
The baseline text normalization algorithms with which the performance of \model{} is compared, are described in Section~\ref{sec:baselines}. 
Next, to check the effectiveness of various text normalization algorithms, we focus on two downstream applications over noisy text -- retrieval and stance detection. The experiments related to these two downstream applications are described in Section~\ref{sec:expt-retrieval} and Section~\ref{sec:stance-detect} respectively. Each of these sections describe the datasets used, the retrieval / stance detection models used, and the experimental results.  
Finally, we conclude the study in Section~\ref{sec:conclusion}.

%and a good text normalizing algorithm can boost the performance of an IR system by retrieving more relevant documents. However, designing a text normalizer for an IR system is very challenging.  Identification of a word and its tenses, disambiguation of word-senses, identification of named entities are few amongst the many challenges the text normalizing system has to tackle.  

\section{Related work}\label{sec:related}
% \vspace{-3mm}

Textual content can contain different types of noise~\cite{Subramaniam-noisy-text-survey}, and
significant amount of research has been done on noisy text.
Some prior works attempted to understand the effect of noise present in text data on applications such as classification~\cite{Vinciarelli-noisy-text-categorize}. 
Also some works have explored the use of {\it artificially generated} noise for cleansing noisy text. For instance, Gadde et al.~\cite{phani-artificial-noise} proposed methods for artificially injecting noise in text, through random character deletion, replacement, phonetic substitution, simulating typing errors, and so on.

Several methods have been developed for the normalization of noisy text, especially microblogs~\cite{Gouws:2011:UML:2140458.2140468,li-liu-2014-improving,Han:2011:LNS:2002472.2002520}.
Based on the approach taken, text normalization methods can be broadly classified into two categories:

\vspace{2mm}
\noindent {\bf Supervised text normalization methods:} Such approaches rely on a training dataset, which is usually a parallel corpus of canonical (correct) and noisy strings. 
For example, in~\cite{brill-moore-2000-improved}, a noisy channel model was employed to perform text normalization.
Incorporation of phonetic information has been shown to help text normalization,
e.g., Toutanova et. al~\cite{toutanova-moore-2002-pronunciation} included word pronunciation information in a noisy channel framework. 
GNU Aspell (\url{http://aspell.net/}), a popular spelling correction tool, also uses phonetic and lexical information to find normalized words. 

More recently, neural models such as RNN~\cite{deepnorm} and Encoder-Decoder models~\cite{lusetti-etal-2018-encoder} have also been used for text normalization. 
Some prior works have also modeled the text normalization problem as machine translation from the noisy text to the clean version~\cite{ling-etal-2013-paraphrasing}.

Supervised methods require huge amounts of training data in the form of a parallel corpus, i.e., pairs of noisy and clean text. 
Such a parallel corpus is often difficult to get, especially for low-resource languages. Even for resource-rich languages, creating a training dataset involves a lot of human intervention, thus making it a costly process. Unsupervised methods try to mitigate some of these problems.

\vspace{2mm}
\noindent{\bf Unsupervised text normalization methods:} These methods make use of the contextual and lexical information available from the corpora to perform normalization, without the need for any training data. 
For instance, Kolak et al.~\cite{Kolak-resnik-2002} applied a noisy channel based pattern recognition approach to perform OCR text correction. 
Ghosh et al.~\cite{ghosh-ipm} used a combination of contextual and syntactic information to find the morphological variants of a word. 
Word embeddings (e.g., Word2vec~\cite{Mikolov2013}) have shown great potential in capturing the contextual information of a word, and hence
word embedding based contextual information has been used in many of the recent works~\cite{rangarajan-sridhar-2015-unsupervised,bertaglia2016exploring}. 
For instance, Sridhar~\cite{rangarajan-sridhar-2015-unsupervised} used contextual information to find a set of candidate variants, which is refined based on a lexical similarity score. 
%Some elements are further removed from the set using a lexical similarity based score. Bertaglia et al.~\cite{bertaglia2017exploring} employs a similar approach wherein the candidate set is formed based on the contextual similarity. The removal step, however, is based on both lexical and contextual similarity.   

In this work, we propose an unsupervised text normalization algorithm, and show that it performs competitively with some of the methods mentioned above~\cite{rangarajan-sridhar-2015-unsupervised,ghosh-ipm}.
%While the baseline methods perform well for only a particular type of noise, our proposed method generalizes better to different types of noise (machine-generated and user-generated).

\vspace{2mm}
\noindent{\bf Task-independent vs. Task-specific text normalization methods:}
Text normalization methods can also be categorized into two broad categories based on whether a method is meant for a specific task.
Most of the methods stated above, including spelling correction tools, are task-independent -- they aim to correct the misspelled words in a corpus.
On the other hand, some task-specific text normalization methods have also been proposed.
For instance, Satapathy et al.~\cite{twitter-normalization-sentiment} proposed a method for normalizing tweets specifically for sentiment analysis. 
Vinciarelli et al.~\cite{Vinciarelli-noisy-text-categorize} attempted normalization of noisy text specifically for categorization.
Again, Ghosh et al.~\cite{ghosh-ipm} used co-occurrence counts and edit-distance measures to find morphological variants of query words for the task of information retrieval from OCR-ed documents. 
%Our proposed method in this paper is also a task-specific one, which targets text normalization for information retrieval.

The proposed method is a task-independent one, and the experiments in the paper show that the data cleansing performed by the proposed method can be helpful for various tasks such as retrieval and stance detection.

%\vspace{2mm}
%\noindent{\bf Simulating and understanding the effects of noise in text data:}

%More recently, the NLP Augmentation library (\url{https://github.com/makcedward/nlpaug}) has been developed for simulating various types of noise in text, including typing errors, OCR errors, as well as random character deletions, substitutions, and so on~\cite{ma2019nlpaug}.

\section{Proposed Text Normalization Algorithm}\label{sec:approach}
% \vspace{-3mm}

This section describes the proposed unsupervised text normalization algorithm, which we call \model{}. 
The implementation is publicly available at \url{https://github.com/ranarag/UnsupClean}.
We start this section with an overview of our proposed algorithm, and then describe every step in detail.

\subsection{Overview of our algorithm}

We look at text normalization from the perspective of identifying {\it morphological variations} of words. To understand this objective, we first need to understand what a {\it morpheme} is. 
According to Morphology (the area of linguistics concerned with the internal structure of words), a  morpheme is the smallest unit of a word with a meaning.\footnote{\url{https://www.cs.bham.ac.uk/~pjh/sem1a5/pt2/pt2_intro_morphology.html}}
Structural variations to a morpheme or a combination of morphemes are known as morphological variations. There can be various types of morphological variations, including the plural and possessive forms of nouns, the past tense, past participle and progressive forms of verbs, and so on. Our approach attempts to identify such (and other) morphological variants of words (which are combinations of morphemes).

%Our approach is based on two ideas -- (i)~string similarity and (ii)~contextual similarity among words.
Intuitively, most morphological variations of a word have high {\it string similarity} with the canonical form of the word. So a string similarity based approach can be used to find candidate morphological variants of a word. 
However, string similarities can be misleading, e.g., the words `{\it Kashmiri}' and `{\it Kashmira}' have high string similarity, but one is an adjective (meaning `belonging to the state of Kashmir, India') while the other is the name of an Indian actress. 
In such cases, it is useful to check the {\it contextual similarity} of the words. 

An important question is how to measure contextual similarity between words.
One potential way is to check {\it co-occurrence counts of the words}, i.e., how frequently the words appear in the same document (same context). 
%that can give the information of how likely two words are to appear in the same context. 
However, for short-length documents (e.g., microblogs), the co-occurrence counts will mostly be zero (or very low), since two variants of the same word would usually not occur in the same microblog. 
%Hence a different method must be applied to such cases. 
An alternative method is to utilize word embeddings that capture contextual features of a word (e.g., Word2vec~\cite{Mikolov2013}).
Hence, our algorithm uses a mixture of string similarity, word co-occurrence counts, and embeddings to find the morphological variants of words.

%Our proposed model considers as input (i)~a corpus $C$ of noisy documents, and (ii)~a query $Q$ (for which retrieval is to be performed).
%We consider $Q$ to be a collection of words $Q = \{ w_i \}_{i=1}^{q}$, where $q$ is the number of words in the query.
%For each word $w \in Q$, our model finds a set $V(w)$ of morphological variants of $w$, where every element of $V(w)$ is a term in the corpus $C$. 
%Thus, an expanded query $Q^{exp}$ is constructed, containing the set of morphological variants of each $w \in Q$, i.e., $Q^{exp} = \bigcup_{i=1}^{q} V(w_i)$.
%This expanded query $Q^{exp}$ is used for retrieval.

\begin{table}[t]
    %\vspace{-10pt}
    \scriptsize
        \centering
        \addtolength{\tabcolsep}{-1pt}
        \begin{tabular}{|l|p{0.35\columnwidth}||l|p{0.35\columnwidth}|}
            \hline
    %		\textbf{Notation}	& \textbf{Description} & \textbf{Notation}	& \textbf{Description}\\ 
              {\bf Symbol}  & \textbf{Description} & 	{\bf Symbol} & \textbf{Description}\\ 
            \hline
            $C$ & Corpus of documents which needs to be cleaned &  & \\
            \hline
            $L$ & Lexicon of noisy words & $L_c$ & Lexicon of clean words  \\
            \hline
            $w_L$ & Noisy word & $w_{L_c}$ & Clean word \\
            \hline
            
            $Cl_{w_{L_c}}$ & cluster of morphological variants of word $w_{L_c}$ &  $CL$ & set of clusters  \\
            \hline

            $BLCS(w_1, w_2)$ & Bi-character Longest Common Subsequence between $w_1$ and $w_2$ & $LCSR(w_1, w_2)$ & Longest Common Subsequence Ratio between words $w_1$ and $w_2$ \\
            \hline

            $BLCSR(w_1, w_2)$ & $BLCS$ Ratio between $w_1$ and $w_2$ & $\alpha$ & Threshold used to form cluster of words having $BLCSR$  \\
            \hline
            
            $ES(w_1, w_2)$ & Edit Similarity between words $w_1$ and $w_2$ & $A(w_{L_c})$ &  cluster of words having lexical similarity with $w_{L_c}$ > $\alpha$  \\
            \hline
            
            $G(w)$ & Graph of words as nodes and edges weighted with the contextual similarity amongst them &  $\beta$ & a threshold used to prune the graph $G$ \\
            \hline            
            $\gamma$ & threshold used in Ghosh et. al.~\cite{ghosh-ipm} & $n$ & threshold used in Enelvo~\cite{bertaglia2016exploring} \\
            \hline	

        \end{tabular}
        \caption{{\bf Notations used in this paper.}}
        \label{tab:notations}
        \vspace{-5mm}
      \end{table}

Table~\ref{tab:notations} lists the notations used in this paper. 
Our proposed model considers as input a corpus $C$ of noisy documents (that need to be cleaned). 
Let $L$ be the set of all distinct words contained in all the documents in $C$. 
Let $L_c$ be a lexicon of words with correct spellings, and let $w_{L_c} \in L_c$ be a correct word.
For each word $w_{L_c} \in L_c$, our model finds a set $Cl_{w_{L_c}}$ of morphological variants of $w_{L_c}$, where every element of $Cl_{w_{L_c}}$ is a term in $L$. 
To find the set $Cl_{w_{L_c}}$ of morphological variants of $w_{L_c}$, our model employs the following four steps: 
(1)~First, we use structural/lexical similarity measures to generate a set of  candidate morphological variants for $w_{L_c}$. 
(2)~Then we construct a similarity graph of the morphological variants, based on the similarity between word embeddings and co-occurrence counts of words.
%This graph is used to identify contextual clusters of morphemes.
(3)~We fragment the graph into clusters of similar morphological variants, using a graph clustering algorithm~\cite{fortunato-comm-detection}. 
This step forms cluster of words which have  both high semantic similarity as well as syntactic similarity with $w_{L_c}$. (4)~Lastly, we choose that cluster of morphological variants as $Cl_{w_{L_c}}$, which has the member with the highest structural similarity with $w_{L_c}$. 
We then, club $Cl_{w_j}$ to $Cl_{w_{L_c}}$ for all $w_j \in Cl_{w_j}$ belonging to $Cl_{w_{L_c}}$. 
Subsequently, all words in $Cl_{w_{L_c}}$ are replaced by $w_{L_c}$.
This results in non-overlapping clusters of morphological variants. Finally, all words belonging  to a particular cluster are replaced by the same word.

\begin{figure}[t]
    \centering
    \includegraphics[scale=0.3]{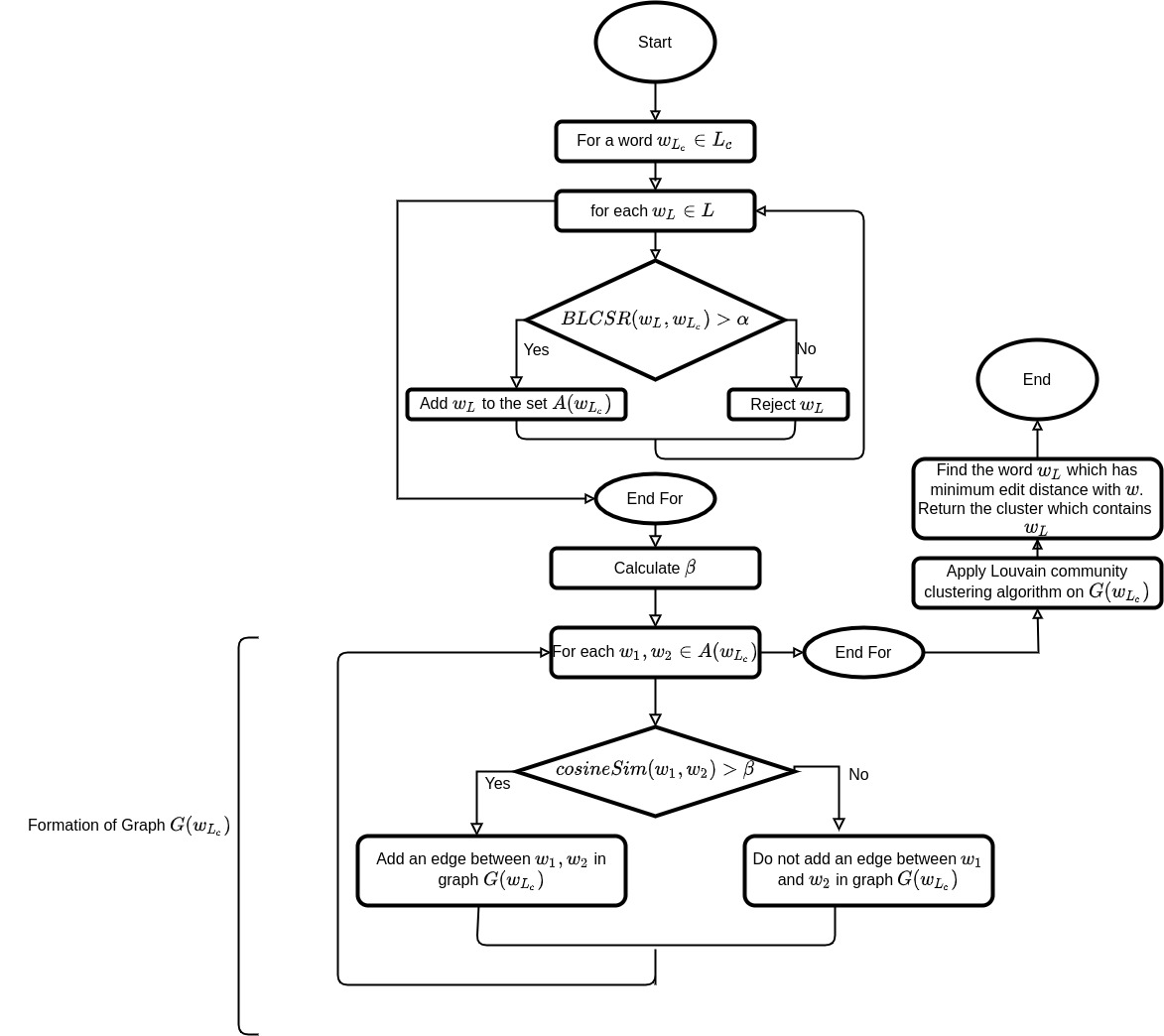}
    \caption{An overview of the proposed algorithm \model{}}
    \label{fig:embedding_model}
    % \vspace{-5mm}
    \end{figure}

Figure~\ref{fig:embedding_model} shows an overview of our proposed algorithm \model{}. 
We explain each of the above steps in the following subsections.

\subsection{Step 1: Generating a set of candidate morphological variants}
\label{sub:algo-generate-candidate-morphemes}

For a given word $w_{L_c}$ (that is considered to be taken from the lexicon of correct words $L_c$), we first construct a set of morphological variants $A(w_{L_c})$, that would include all possible noisy variants of $w_{L_c}$ that occur in the given corpus.
%We analyzed different forms of noise in different sources of textual data, including social media data, OCR-ed text, and so on. 
%Textual data from social media sites like Facebook, Twitter and also from SMS and chat messages broadly have a lot of out-of-vocabulary (OOV) words which can be broadly classified into three categories:
We observed that various noisy variants of words %appearing in noisy texts (OCR-ed text, social media posts, etc.) 
can be broadly classified into the following categories:\\ 
(1)~Vowels are often added/deleted from words, especially in social media posts. E.g., `{\it cool}' is written as `{\it cooool}', and `{\it thanks}' is written as `{\it thnks}'.\\
(2)~Platforms like microblogging sites (e.g., Twitter, Weibo) allow very short posts, with a hard limit on the number of characters. On such platforms, users shorten words arbitrarily, omitting not only vowels but also many consonants. For instance, `{\it tomorrow}' is shortened as `{\it tmrw}', and `{\it medicines}' as `{\it meds}'. Even named entities can be shortened, e.g., `{\it Kathmandu}' as `{\it ktm}'~\cite{roy-cikm17}.\\
%\item {\it Existence of code-mixed documents.} While chatting or posting in social media people sometimes put colloquial terms in their texts. Such terms can be from a language different from the language they are texting in. For example a person tweeted  after watching a movie "@JazbaaFilm ....love this movie and love irrfan's fadu dialogues....awesome" in order to express his/her positive sentiment towards the movie, where {\it fadu} is a term that has its existence in the hindi language and not english.
(3)~People sometimes misspell words due to the phonetic similarity between the correct word and the misspelled word. For example, the phrase `smell like men's cologne' can be misspelled as `smell like men's colon'.\\
%Though the word `{\it colon}' is a correct English word (existing in an English dictionary), it is not correct in the given context.
(4)~Different from the types of user-generated noise described above, there is also machine-generated noise (e.g., OCR noise) where spelling errors can occur in any part of the word (including the first character), whereas in case of human-generated variants, the first few characters usually match with that of the correct word. 
Also in case of OCR noise, there is a high chance of {\it not} having the correct word at all in the corpus, while a corpus containing user-generated errors usually contains both the clean and noisy variants of words. 
%The challenge involving the first kind of noise can be handled by removing vowels from words \myself{Add citations}. However, this method has two problems:
%\begin{itemize}
%    \item one needs to identify the language of the corpus which is difficult in case of social media documents which have a lot of code-mixing in them
%    \item it introduces some noise in the corpus. For example words like {\it leave} and {\it live} get truncated to {\it lv}.
%\end{itemize}

To capture all these different types of noisy variants, we include in $A(w_{L_c})$, every word that has {\it Bi-character Longest Common Sub-sequence Ratio (BLCSR) similarity} with $w$ greater than a threshold $\alpha \in (0, 1)$. Formally: 
\begin{equation}
    A(w_{L_c}) = \{w_L \mid \Call{BLCSR}{w_L,w_{L_c}} > \alpha \}
\end{equation}
where $BLCSR$ is the Bi-character Longest Common Sub-sequence Ratio~\cite{melamed-1995-automatic} based string similarity measure between two strings, and $\alpha \in (0, 1)$ is a hyper-parameter. $BLCSR$ is $LCSR$ with bi-gram of characters calculated as:
\begin{equation}
        BLCSR(w_1, w_2) = \frac{BLCS(w_1, w_2)}{max length(w_1, w_2) - 1.0}
\end{equation}
where $BLCS(w_1, w_2)$ is the Longest Common Subsequence considering bi-gram of characters.
For example, if $w_1$ is `ABCD' and $w_2$ is `ACD' then its $BLCS$ value will be 1. 
Once $A(w_{L_c})$ is formed, we remove the word $w_{L_c}$ itself from $A(w_{L_c})$.

\subsection{Step 2: Constructing a similarity graph of morphological variants}
\label{sub:algo-construct-sim-graph}

As stated earlier in this section, only string similarity based measure is not sufficient for capturing the morphological variations of words. 
%For example {\it quite} and {\it quit} are two very different words with very different meaning but having a high string similarity between them. 
%Hence, we need to remove such words from the set by using contextual information. 
Hence, we also consider the contextual similarity between words. 
We employ two ways of measuring contextual similarity between words -- 
(i)~observing co-occurrence counts. which works better for longer documents (that are more likely to contain multiple variants of the same word), and (ii)~computing similarity between word embeddings which works better for short documents such as microblogs (since embeddings tend to lose the context over long windows).
%We have observed that co-occurrence usually works better than word embeddings in cases of long documents, while cosine similarity of word-embeddings perform better in case of short documents. 
%This is because word-embeddings tend to loose the context over long windows whereas co-occurrence counts have no such problem. 
%On the other hand, different variants of the same word are much more likely to appear in long documents, than in short documents.
%Hence, we use both co-occurrence count and cosine similarity between word-embeddings in this step. 

Let $\overrightarrow{w_i}$ be the embedding of the word $w_i$.
Specifically, we obtain the word embeddings using Word2vec~\cite{word2vec-mikolov} over the given corpus.
Word2vec is a neural network based word-embedding generation model~\cite{word2vec-mikolov}. 
Unlike one-hot vectors for words, word2vec embeddings capture some contextual information in them -- two words having similar word2vec embeddings have been used in similar context in the corpus on which Word2vec has been trained. 
For instance, after training word2vec on the `TREC 2005 confusion' dataset, words like `program`, `programs`, `program1`, and `program:s` have high cosine similarity amongst their embeddings.
Here, a word $w_i$ is considered to be used in a similar context as another word $w_j$ if $w_i$ frequently appears `nearby' $w_j$, i.e., within a fixed context window size around $w_j$.
Another advantage of word2vec embeddings over one-hot vectors is that the size of these embeddings do not increase with increase in vocabulary.
Two different versions of the Word2vec model exist depending upon how the embeddings are generated -- (1)~skip-gram, and (2)~Continuous Bag of Words (CBoW). 
For our algorithm, we used Word2vec with context window size $3$ and skip-gram based approach.
We chose skip-gram over CBoW since it generates better word embeddings in case of infrequent words.\footnote{See \url{https://code.google.com/archive/p/word2vec/}.} 
Also, we kept the context window size 3 so as to ensure good word embeddings for infrequent noisy words.

We construct a graph $G(w_{L_c})$ with all words from the set $A(w_{L_c})$ as the nodes. 
We consider an edge between two nodes (words) $w_i, w_j \in A(w_{L_c})$ only if the value of cosine similarity between the embeddings $\overrightarrow{w_i}$ and $\overrightarrow{w_j}$  is greater than a threshold $\beta$. 
We calculate the $\beta$ value by using a weighted average of the cosine similarities between embeddings of all word-pairs in the set $A(w_{L_c})$, with the BLCSR value between the words as the corresponding weights. 
That is:
\begin{equation}
    \beta = \frac{\sum_{w_i, w_j \in A(w_{L_c})} BLCSR(w_i, w_j) * cosineSim(\overrightarrow{w_i}, \overrightarrow{w_j})}{\sum_{w_i, w_j \in A(w_{L_c})} BLCSR(w_i, w_j)}
    \label{eq:beta}
\end{equation}
This is done to find a threshold which takes both the syntactic similarity and semantic similarity into consideration. 
If $cosineSim(\overrightarrow{w_i}, \overrightarrow{w_j})$ is greater than $\beta$, it means that the word-embedding model is highly confident of $w_i$ being a morphological variant of $w_j$ (and vice versa), and hence we connect the two words with an edge. 
We weight the edge between the two words $w_i, w_j$ with weight $W_{ij}$ defined as:
\begin{equation}
    W_{i,j} = cosineSim(\overrightarrow{w_i}, \overrightarrow{w_j}) \times Cooccur(w_i, w_j)
    \label{eq:edgeweights}
\end{equation}
where $Cooccur(w_i, w_j)$ is the co-occurrence count of the words $w_i$ and $w_j$ in the same document (averaged over all documents in the corpus).
We take the product of the two similarities %(instead of, say, weighted average) 
since cosine-similarity of word embeddings captures similarity of the words in a local context (e.g., in a context window of size 3 words) and co-occurrence count captures their similarity in a global context (of a whole document). 
%For cases where the document sizes are small (e.g. Twitter corpus) the co-occurrence counts will be small for almost all pairs of words and the cosine-similarity will play a major role in determining the strength of edges. 
%Whereas for longer documents, the cooccurrence counts also play an important role in identifying morphological variants.

%After the formation of the graph we fragment the graph to smaller clusters.

\subsection{Step 3: Fragmenting the graph into clusters of morphological variants}
\label{sub:algo-fragment-graph}

In the graph $G(w_{L_c})$, edges exist only between those nodes (words) which have at least a certain amount of similarity ($cosineSim(\overrightarrow{w_i}, \overrightarrow{w_j}) > \beta$). 
Next, we want to identify groups of words which have a high level of similarity  among themselves. 
%In other words, we want to group nodes which have high-weight edges among them. 
To this end, community detection or graph clustering algorithms can be used to identify groups of nodes that have `strong' links (high-weight edges) among them. 
There are many graph clustering algorithms in literature~\cite{fortunato-comm-detection}.
We specifically use the popular Louvain graph clustering algorithm~\cite{louvain-community}.
This algorithm functions on the principle of maximizing a {\it modularity score}~\cite{newman-modularity} for each cluster/community, where the modularity quantifies the quality of an assignment of nodes to clusters, by evaluating how much more densely connected the nodes within a cluster are, compared to how connected they would be in a random network. 
Thus, it can be expected that all words in a particular cluster (as identified by Louvain algorithm) have strong semantic similarity amongst them.

%After the clustering step we move on to find the most suitable clusters of morphological variants to the given query word. 

\subsection{Step 4: Finding the most suitable cluster for a word}
\label{sub:algo-most-suitable-morpheme-cluster}

%In the previous step, we partitioned the graph $G(w)$ into clusters of words that have both high syntactic and contextual similarities amongst them. 
Out of the clusters identified from $G(w_{L_c})$,
we now need to find that  cluster $Cl_{w_{L_c}}$ whose members are most likely to be the morphological variants of the given word $w_{L_c}$. 
We consider $Cl_{w_{L_c}}$ to be that cluster which contains the word with the minimum edit-distance from the word $w_{L_c}$ (ties broken arbitrarily). 
%Note that, for this step, we consider only the string similarity between words (and not their semantic similarity) because we do {\it not} train the word embedding model on the queries along with the whole corpus. This approach makes our approach suitable for online retrieval as well, where unseen queries can be handled.

\vspace{2mm}
\noindent Thus, for a given word $w_{L_c}$, we identify a set $Cl_{w_{L_c}}$ of morphological variants.
Some examples of $Cl$s identified by the proposed algorithm are shown later in Table~\ref{tab:examples-microblog} and Table~\ref{tab:examples-confusion}.

%As stated earlier, these steps are repeated for each word $w_i \in Q$ (where $Q$ is the given query), to obtain an expanded query $Q^{exp} = \bigcup_{i=1}^{q} V(w_i)$. This expanded query is used for retrieval.

\section{Baseline text normalization approaches}\label{sec:baselines}
% \vspace{-3mm}

We compare the performance of our proposed text normalization algorithm with those of several baseline normalization methods that are described in this section. Since the implementations of these baselines are not publicly available (except Aspell and Enelvo), we implemented the algorithms following the descriptions in the respective papers, with some modifications as described below. Code for Enelvo is publicly available but as a normalizer package for Portuguese Language. Hence we implemented the Enelvo algorithm as well.

\subsection{Aspell}

\begin{figure}[t]
    \centering
    \includegraphics[scale=0.4]{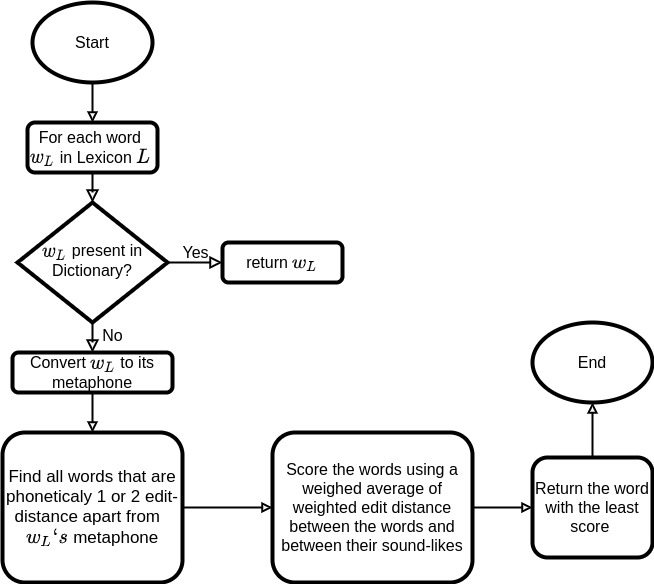}
    \caption{An overview of GNU Aspell}
    \label{fig:aspell}
    % \vspace{-5mm}
    \end{figure}
    
Aspell is an open-source spell checking library which uses a combination of Lawren Philips' metaphone search algorithm~\cite{Philips:2000:DMS:349124.349132} and Ispell's missed strategy mechanism. 
Given a (misspelled) word, the algorithm first finds all the metaphones of the word. It then finds all the words that are within 1 or 2 edit-distance from any of the metaphones of the word. 
Figure~\ref{fig:aspell} shows a flowchart describing the functioning of Aspell. 

We used the GNU implementation of Aspell with the Python wrapper available at \url{https://github.com/WojciechMula/aspell-python}. For the RISOT dataset (in Bengali) that we use later in the paper, we used the Aspell Bengali dictionary available at
\url{https://ftp.gnu.org/gnu/aspell/dict/0index.html}.

%However, edit distance is incapable of handling homonymy (e.g. {\it industrious} and {\it industrial}).
% \vspace{-5pt}

\subsection{Sridhar}

\begin{figure}[t]
    \centering
    \includegraphics[scale=0.4]{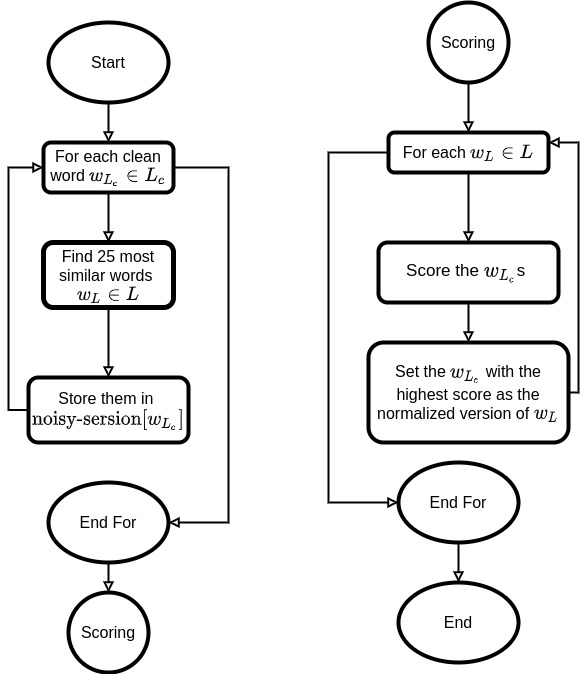}
    \caption{The overview of the algorithms of both Sridhar~\cite{rangarajan-sridhar-2015-unsupervised} and Enelvo~\cite{bertaglia2016exploring}. The only difference between the  two algorithms
    is the way in which the scores are calculated.}
    \label{fig:sridhar-enelvo}
    % \vspace{-5mm}
\end{figure}

%Sridhar~\cite{rangarajan-sridhar-2015-unsupervised} proposed an approach which unlike any spell-checker model like Aspell, involves contextual information along with it the syntactical information to perform text normalization. 
Unlike a spell-checker like Aspell, the approach by Sridhar~\cite{rangarajan-sridhar-2015-unsupervised} uses both contextual information and structural information for text normalization. Figure~\ref{fig:sridhar-enelvo} gives a schematic flowchart describing the method.

The approach involves a lexicon $L_c$ of correct words (i.e. $w_{L_c} \in L_c$) such that there exists a word embedding for each $w_{L_c}$. Noisy versions for each $w_{L_c}$ are found from the corpus as:
\begin{equation}
    \mbox{noisy-version}[w_{L_c}] = \displaystyle\forall_{w_L \in L, w_L \notin L_c} \, \max_k \mbox{cosineSim}(\overrightarrow{w_{L_c}}, \overrightarrow{w_L})
    \label{eqn:sridhar-step1}
\end{equation}
where $w_L$ is a noisy word from the corpus $C$ with lexicon $L$. 
For each $w_{L_c}$, $k=25$ nearest neighbors are found and stored in the map noisy-version. 
Note that only contextual similarity among word embeddings is considered for identifying noisy versions in this initial step.

Now for a given noisy word $w_L \in L$, a list of all those $w_{L_c}$ is obtained for which noisy-version[$w_{L_c}$] contains $w_L$. 
The words $w_{L_c}$ in this list are scored, where the score is calculated as: 
\begin{equation}
    \mbox{score}(w_{L_c}, w_L)=\frac{LCSR(w_L, w_{L_c})}{ED(w_L, w_{L_c})}
\end{equation}
where $ED(w_L, w_{L_c})$ is the edit-distance between $w_L$ and $w_{L_c}$, and $LCSR(w_L, w_{L_c})$ is the Longest Common Subsequence Ratio~\cite{melamed-1995-automatic} calculated as
%$LCSR(w_1, w_2) = \frac{LCS(w_1, w_2)}{max length(w_1, w_2)}$,
\begin{equation}
    LCSR(w_L, w_{L_c}) = \frac{LCS(w_L, w_{Lc})}{max length(w_L, w_{L_c})}
\end{equation}
where $LCS(w_L, w_{L_c})$ is the {\it Longest Common Subsequence} Length between the words $w_L$ and $w_{L_c}$.

Finally, the $w_{L_c}$ with the highest score (as computed above) is considered to be the normalized version of the noisy word $w_L$.
Note that the computation of {\it score} considers only the lexical similarity between words. The integration of contextual similarity with lexical similarity is achieved by first selecting the $w_{L_c}$ words based on contextual similarity with $w_L$, and then computing the lexical similarity score.

In the original paper~\cite{rangarajan-sridhar-2015-unsupervised}, the author created $L_c$ (the lexicon of correct words) by training a word embedding model on 
some randomly selected sentences from the corpus which were {\it manually normalized by a professional transcriber}. 
Since our objective is to {\it not} use human resource, we chose pre-trained word-vectors to construct $L_c$. 
% \todo{check if the symbol will be $L_c$; it was earlier written $C$}
Specifically, for the RISOT dataset (in Bengali), we used pre-trained word2vec word vectors on Bengali Wikipedia data (available at  \url{https://github.com/Kyubyong/wordvectors}), and for the English / microblog datasets, we used the pre-trained Word2Vec word vectors on the English Google News data (available at \url{https://github.com/eyaler/word2vec-slim}).

\subsection{Enelvo}
%Bertaglia et. al.~\cite{bertaglia2016exploring} developed a way to normalize Portuguese user-generated content. 
%Unlike Aspell which uses only syntactic and metaphonic information to find the correct variant, this method uses {\it contextual information} which it captures using word embeddings. 
This method~\cite{bertaglia2016exploring} is inherently similar to that of Sridhar~\cite{rangarajan-sridhar-2015-unsupervised}. Figure~\ref{fig:sridhar-enelvo} gives a schematic flowchart describing both these methods.
Enelvo involves a lexicon $L_c$ of correct words (i.e., $w_{L_c} \in L_c$) with frequency $\geq 100$ in the given corpus. The noisy versions $w_L \in L$ (the lexicon the given corpus) of the word $w_{L_c}$ are then found as:
\begin{equation}
    \mbox{noisy-version}[w_{L_c}] = \displaystyle\forall_{w_L \in L, w_L \notin L_c} \, \max_k \mbox{cosineSim}(\overrightarrow{w_{L_c}}, \overrightarrow{w_L})
    \label{eq:enelvo-knn}
\end{equation}
where $w_{L_c}$ is the correct word from the lexicon $L_c$, and $w_L$ is the noisy word from the lexicon $L$ of corpus $C$. 
For each $w_{L_c}$, the top $k=25$ most similar noisy words $w_L \in L$ are found and stored in the map noisy-version.
This step is similar to the first step in the Sridhar baseline~\cite{rangarajan-sridhar-2015-unsupervised}, and considers only contextual similarity among word embeddings.

Next, for a given noisy word $w_L \in L$ a list of all those $w_{L_c}$ is chosen for which noisy-version[$w_{L_c}$] contains $w_L$. The words $w_{L_c}$ in this list are then scored. 

The score is computed as:
\begin{equation}
    score(w_L, w_{L_c}) = n \times \mbox{lexical-similarity}(w_L, w_{L_c}) + (1 - n) \times \mbox{cosineSim}(\overrightarrow{w_L}, \overrightarrow{w_{L_c}})
    \label{eq:enelvo-score}
\end{equation}
where $n \in (0, 1)$ is a hyperparameter,
%cosineSim($\overrightarrow{w}, \overrightarrow{w_c}$), is the cosine similarity between the embeddings of $w$ and $w_c$, 
and lexical-similarity($w_L, w_{L_c}$) is measured as:
\begin{equation}
    \mbox{lexical-similarity}(w_L, w_{L_c})= \begin{cases}
                                    \frac{LCSR(w_L, w_{L_c})}{MED(w_L, w_{L_c})}, & \text{if $MED(w_L, w_{L_c}) > 0$}.\\
                                    LCSR(w_L, w_{L_c}), & \text{otherwise}.
  \end{cases}
\end{equation}
Here $MED(w_L, w_{L_c}) = ED(w_L, w_{L_c}) - DS(w_L, w_{L_c})$ is the modified edit distance between $w_L$ and $w_{L_c}$, where
%calculated as $MED(w_L, w_{L_c}) = ED(w_L, w_{L_c}) - DS(w_L, w_{L_c})$, 
$ED(w_L, w_{L_c})$ is the edit distance and $DS(w_L, w_{L_c})$ is the {\it diacritical symmetry} between $w_L$ and $w_{L_c}$ (a feature that is useful in Portuguese and some non-English languages -- see~\cite{bertaglia2016exploring} for details). 
$LCSR(w_L, w_{L_c})$ is the {\it Longest Common Subsequence Ratio}, calculated as:
\begin{equation}
    LCSR(w_L, w_{L_c}) = \frac{LCS(w_L, w_{L_c}) + DS(w_L, w_{L_c})}{max length(w_L, w_{L_c})}
\end{equation}
%$LCSR(w_L, w_{L_c}) = \frac{LCS(w_L, w_{L_c}) + DS(w_L, w_{L_c})}{max length(w_L, w_{L_c})}$
where $LCS(w_L, w_{L_c})$ is the {\it Longest Common Subsequence} of words $w_L$ and $w_{L_c}$.
Hence, the computation of {\it score} integrates both lexical similarity as well as contextual similarity, which is a point of difference from the Sridhar baseline~\cite{rangarajan-sridhar-2015-unsupervised}.
Finally, the $w_{L_c}$ with the highest score is chosen as the normalized version of $w_L$.

Thus, the integration of contextual similarity and lexical similarity is achieved in Enelvo as follows -- first, the top $k$ noisy words $w_L$ are chosen for each clean word $w_{L_c}$ using contextual similarity. Then for a given noisy word $w_L$, the best clean word $w_{L_c}$ is selected using a scoring function that is a weighted combination of contextual similarity and lexical similarity.

For the embeddings, the authors of~\cite{bertaglia2016exploring} trained a Word2Vec skip-gram model on a conglomeration of two corpora -- Twitter posts written in Portuguese and a product review corpus containing both noisy and correct texts~\cite{hartmann-etal-2014-large}. 
Since our datasets are in English and Bengali, we used 
%Since our retrieval tasks do not involve Portuguese documents and also due to the inavailability of the Twitter Portuguese corpus we used  
pre-trained embeddings for the correct words $w_{L_c} \in L_c$ and trained word2vec models on the respective noisy corpora for the noisy word embeddings $w_L \in L$. 
For the embeddings of $w_{L_c}$ in the RISOT dataset, we used pretrained Word2Vec embeddings of the Bengali Wikipedia dataset (available at  \url{https://github.com/Kyubyong/wordvectors}). 
Similarly, for the English and microblog datasets, we used pretrained word embeddings of the Google English News dataset (available at \url{https://github.com/eyaler/word2vec-slim}).

\subsection{Ghosh} 
\begin{figure}[t]
    \centering
    \includegraphics[scale=0.4]{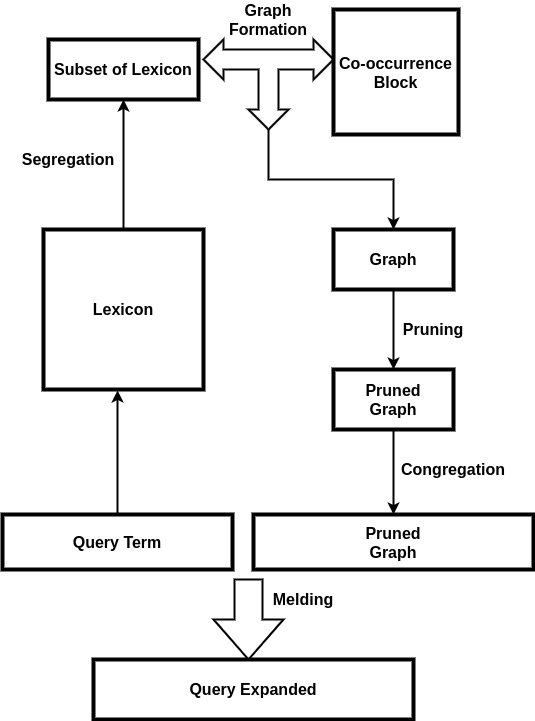}
    \caption{An overview of the algorithm by Ghosh et. al., derived from the original paper~\cite{ghosh-ipm}}
    \label{fig:ghosh_model}
    \end{figure}

Ghosh et. al~\cite{ghosh-ipm} employs a five-step approach to find the morphological variants of a word (see Figure~\ref{fig:ghosh_model}).
The first step called {\it segregation} involves removal of words with string similarity $\leq \alpha$ (where $\alpha \in (0, 1]$) with the correct word $w_{L_c}$(or query word), resulting in the formation of $A(w_{L_c})$. The next step is {\it Graph Formation} where a graph $G=(V,E)$ is formed with vertices $V$ as the words in $A(w_{L_c})$ and edges $E$ with weights as the co-occurrence counts between them. Next step is the {\it Pruning} step where the Graph $G$ is pruned. Pruning is carried out only if the maximum edge-weight ($max_{ew}$) in $G$ is greater than a threshold $\gamma$ which is determined by the user.
In this step, edges which are less than $\beta \; \%$ of $max_{ew}$ are removed from the Graph $G$ resulting in the formation of a graph $G_r = (V, E')$ where $E' \subset E$. 

The next step is called {\it Congregation} where the graph $G_r$ is further fragmented into clusters based on their edge-weights. Two vertices $v_1$ and $v_2$ belong to the same cluster if they  satisfy any of the following conditions -- (i)~$v_1$ is the strongest neighbour of $v_2$, or (ii)~$v_2$ is the strongest neighbour of $v_1$.
This results in the formation of a set of $k$ clusters $CL = \{Cl_1, Cl_2, \ldots, Cl_k\}$. Congregation step is followed by the final step called {\it Melding}. 
In this step, a cluster is chosen from the set of clusters $CL$ to have the morphological variants of the query word $w_{L_c}$. Such a cluster containing lexically the most similar word to $w_{L_c}$ is chosen to be the cluster containing morphological variants of $w_{L_c}$. The  lexical similarity is measured using {\it Edit Similarity} (ES), which is an edit-distance based similarity metric, having the formula:
\begin{equation}
    ES(w_1, w_2) = 1 - \frac{ED(w_1, w_2)}{max(stringLength(w_1), stringLength(w_2))}
\end{equation}
where $ED(w_1, w_2)$ is the edit-distance between the two words $w_1$ and $w_2$.

\vspace{3mm}
\noindent {\bf Difference of proposed algorithm with baselines:} The baseline normalization algorithms discussed in this section are significantly different from the proposed algorithm \model{}. Sridhar and Enelvo both use contextual and lexical similarity, as does \model{}; however, the approach for integrating contextual similarity and lexical similarity is different in \model{} as compared to Sridhar and Enelvo. Additionally, neither Sridhar nor Enelvo are graph-based methods, whereas \model{} relies on forming a similarity graph and then clustering the graph.
The Ghosh baseline is most similar to \model{} -- both use contextual as well as lexical similarity, and both are graph-based methods. But importantly, Ghosh measures contextual similarity using only co-occurrence statistics, whereas \model{} uses a combination of co-occurrence statistics and word embeddings. The advantage of using word embeddings will be clear from our experiments described in the subsequent sections.

\vspace{3mm}
\noindent
In the next two sections, we compare the performance of the proposed \model{} text normalization algorithm with that of the baseline normalization algorithms described in this section. 
To this end, we consider two downstream applications -- retrieval (Section~\ref{sec:expt-retrieval}) and stance detection (Section~\ref{sec:stance-detect}).

\section{Downstream Application 1: Retrieval}
\label{sec:expt-retrieval}
In this section, we evaluate the performance of a text normalization algorithm by observing the performance of a retrieval (search) system over a noisy corpus that is normalized by the algorithm. 

%We use the proposed \model{}, and the baselines Sridhar, Enelvo and Ghosh as follows.
%For a given query $Q$, we construct an expanded query $Q^{exp}$ by adding the morphological variants for each word in $Q$ (as identified by a text normalization algorithm from the given corpus). The expanded query is used for retrieval using Indri. 
%Specifically, we used the {\it syn} operator of Indri (meant for clubbing the synonyms of a word) to club each query word with its morphological variants. 
%However, since Aspell is a spell correction toolkit,
%retrieval using Aspell is carried out differently, as compared to the other normalization algorithms. 
%For the other algorithms, the original corpus is indexed using Indri, and the expanded query is used for retrieval from the indexed corpus (as explained above). 
%On the other hand, for Aspell, 
%the corpus is first normalized using Aspell, and the normalized corpus is indexed using Indri. The queries are also normalized using Aspell, and then the retrieval is carried out.
%Note that both approaches are comparable with each other. 

\subsection{Datasets for retrieval}
\label{sub:data-retrieval}

For evaluation of various text normalization methods, we experimented with the following three datasets, each having a different source of noise. The statistics of the datasets are given in Table~\ref{tab:data-retrieval}.

\vspace{2mm}
\noindent {\bf (1) RISOT Bengali OCR Data:} This dataset was released in the FIRE RISOT track (\url{https://www.isical.ac.in/~fire/risot/risot2012.html}). The text version of 62,825 news articles of a premiere Bengali newspaper Anandabazar Patrika (2004-2006) was scanned and OCRed to create the collection.
The dataset also contains 66 queries, and their set of relevant documents (gold standard). 

\vspace{2mm}
\noindent {\bf (2) TREC 2005 Confusion Track Data:} This dataset was created for the TREC Confusion track 2005~\cite{Kantor-confusion-2005}. The dataset consists of the text versions of Federal Register English documents (of the year 1994), which were scanned and OCR-ed. 
%having approximate error rates of 5\% and 20\%. 
We use the 5\% version for our experiments, where the approximate character error rate is 5\%. Further details can be found in~\cite{Kantor-confusion-2005}.

\begin{table}[tb]
    \centering
    \footnotesize
    \begin{tabular}{|c|c|c|}
    \hline
    {\bf Dataset}     &  {\bf No. of documents} & {\bf No. of queries}\\
    \hline
     FIRE RISOT Bengali  & 62,825 & 66\\
     TREC Confusion 2005 & 55,600 & 49\\
     TREC Microblog 2011 & 16,074,238 & 50\\
     \hline
    \end{tabular}
    \caption{{\bf Description of the retrieval datasets used for empirical comparison of the proposed and baseline text normalization methods.}}
    \label{tab:data-retrieval}
\end{table}

\vspace{2mm}
\noindent {\bf (3) TREC 2011 Microblog Track Data:} 
%We also look test the ability of our proposed algorithm to adapt to different kinds of noise. To this end, 
We consider an English tweet collection viz. TREC Microblog track collection 2011 (\url{https://trec.nist.gov/data/microblog2011.html}, containing noise produced by cavalier human writing style. 
Unlike the other two datasets, this is a much larger collection (containing over 16 million tweets), and is also expected to test the scalability of any text normalization method.

\vspace{2mm}
\noindent Thus, we use three noisy collections having different types of errors (OCR errors, errors due to users' writing styles) in two different languages (English and Bengali) to test the efficacy and language independence of various text normalization methods.
For brevity, we refer to the datasets as {\bf RISOT}, {\bf Confusion} and {\bf Microblog} respectively.

\subsection{Retrieval Model used}

We used the popular Information Retrieval system {\it Indri} which is a component of the {\it Lemur Toolkit} (\url{https://www.lemurproject.org/indri/}) for the retrieval experiments.
Indri is a language-independent retrieval system, that combines language modeling and inference network approaches~\cite{indri-paper}. 
We used the default setting of Indri for all experiments. 

A particular corpus is first normalized using a normalization algorithm (either \model{}, or one of the baselines Sridhar, Enelvo, Ghosh, and Aspell), and the normalized corpus is then indexed using Indri.
The queries are also normalized using the same normalization algorithm, and then the retrieval is carried out.

\subsection{Parameter settings of normalization algorithms} \label{sub:retrieval-parameter-setting}

The proposed algorithm has one hyperparameter $\alpha$ (the similarity threshold for BLCSR). We obtained the best value for $\alpha$ via grid search in the range $[0.5, 1.0]$ with a step size of $0.01$.
The best values are $\alpha = 0.56$ for RISOT, $\alpha = 0.9$ for Confusion and $\alpha = 0.83$ for the Microblog dataset.

Ghosh et al.~\cite{ghosh-ipm} has three hyper-parameters, namely $\alpha, \beta$ and $\gamma$. We followed the same methodology as stated in the original paper~\cite{ghosh-ipm} to set the parameters -- we varied the $\alpha$ and $\beta$ while keeping the $\gamma$ value equal to $50$.
We obtained the best values for $\alpha$ and $\beta$ via grid search in the range  $[0.5, 1.0]$ with a step size of $0.01$. The best values are $\alpha = 0.8, \beta = 0.7$ for RISOT, $\alpha = 0.9, \beta = 0.7$ for TREC 2011 microblog and $\alpha = 0.7, \beta = 0.8$ for the Confusion dataset.

For both Sridhar~\cite{rangarajan-sridhar-2015-unsupervised} and Enelvo~\cite{bertaglia2016exploring} have a hyper-parameter $K$ (the number of nearest neighbors to a particular word $w$, that are considered as potential variants of $w$).

Both these works directly specify $k = 25$ without any tuning. Hence, we adhered to the value $k = 25$ value provided by the authors in their papers. 
Enelvo has another hyper-parameter $n$ that specifies the relative significance of contextual similarity and lexical similarity while computing {\it score}; we used $n=0.8$, the value used by the authors in the original paper.

\subsection{\bf Evaluation metrics for retrieval}

As stated earlier, to evaluate the performance of a text normalization algorithm, we measure the performance of retrieval on noisy datasets (described in Section~\ref{sub:data-retrieval}) cleaned using the normalization algorithm. 
We use some standard metrics to evaluate the retrieval performance, as follows.

For the RISOT and Microblog datasets, we measure Recall@100 (at rank 100), Mean Average Precision mAP@100 and mAP@1000 (at ranks 100 and 1000) as the metrics for comparing the retrieval performance. 

The TREC 2005 Confusion dataset, however, has only {\it one relevant document per query}. Hence, it is more common to report the Mean Reciprocal Rank (MRR) metric for this dataset.

\subsection{\bf Results}

Table~\ref{tab:result-risot} compares the retrieval performance using different text normalization algorithms (the proposed method and the baselines) for the RISOT Bengali dataset. 
Also shown are the retrieval performances over the raw data, i.e., without any normalization.
Similarly, Table~\ref{tab:result-microblog} shows the results for the TREC 2011 Microblog dataset, while
Table~\ref{tab:result-confusion} shows results for the TREC 2005 Confusion dataset.
%Across all datasets, the proposed model enables better retrieval in general, as compared to the baseline normalization methods. 
The tables also report results of statistical significance testing -- the symbols A, E, G, S in the superscripts in the tables indicate that the proposed method is statistically significantly better at 95\% confidence interval ($p < 0.05$) than Aspell, Enelvo, Ghosh et al., and Sridhar respectively.

For the Microblog dataset (see Table~\ref{tab:result-microblog}), the proposed \model{} performs competitively with the baselines. Retrieval using the Sridhar baseline achieves higher  Recall@100, while retrieval using \model{} achieves higher mAP@100 and mAP@1000 scores. None of the differences are statistically significant.

For the other two datasets which primarily contain OCR noise (see Table~\ref{tab:result-risot} and Table~\ref{tab:result-confusion}), \model{} enables much better retrieval than the baselines.
Especially for the RISOT dataset (in Bengali language), the baselines Enelvo, Sridhar and Aspell perform very poorly. 
In fact, in most cases, retrieval performance is better over the raw RISOT data, than after normalization by these three baselines.
In contrast, \model{} leads to significant improvement in retrieval performance.
Hence, {\it the proposed model can generalize to different types of noise as well as different languages, better than most of the baselines}.

\begin{table}[tb]
    \footnotesize
    \center
    \addtolength{\tabcolsep}{2pt}
    \begin{tabular}{|l|c|c|c|}
        \hline
        \textbf{Normalization Algorithm} & \textbf{Recall@100} &\textbf{mAP@100} &\textbf{mAP@1000} \\ 
        \hline
        \hline
        Raw data (without any cleaning)  & 50.08\% & 17.21\% &18.50\% \\
        \hline \hline
        Ghosh et al.\cite{ghosh-ipm}  & 54.48\% & 19.48\%  & 20.98\% \\
        % total map is 22.14
        \hline
        Enelvo~\cite{bertaglia2016exploring}  & 49.24\% & 17.04\%  & 18.37\% \\
        \hline
        Sridhar \cite{rangarajan-sridhar-2015-unsupervised}  & 47.75\% & 17.38\%  & 18.76\% \\
        \hline
        Aspell  & 24.40\% & 5.43\%  & 5.58\% \\
        \hline
        \hline
        \model{} (Proposed)  & {\bf 58.32\%}$^{ESA}$ & {\bf 21.50\%}$^{ESA}$  & {\bf 23.00\%}$^{ESA}$ \\
        %Significantly better than  & E,S,A  & E,S,A & E,S,A \\
        \hline
    \end{tabular}
        \caption{{\bf Retrieval performance on RISOT Bengali OCR dataset, using Indri IR system. Best performances are in bold font. 
        %The proposed text normalization algorithm enables the best retrieval, compared to all baselines. 
        Superscripts E,S,A indicate the proposed method is statistically significantly better than Enelvo, Sridhar and Aspell respectively.}}
    \label{tab:result-risot}
    % \vspace{-10mm}    
\end{table}

\begin{table}[tb]
    \footnotesize
    \center
    \addtolength{\tabcolsep}{3pt}
    \begin{tabular}{|l|c|c|c|}
        \hline
        \textbf{Normalization Algorithm} & \textbf{Recall@100} &\textbf{mAP@100} &\textbf{mAP@1000} \\ 
        \hline
        \hline
        Raw data (without any cleaning)  & 37.63\% & 16.13\%  & 19.77\%\\
        \hline \hline
        Ghosh et al.\cite{ghosh-ipm} & 37.62\% & 15.97\%  & 19.66\% \\
        \hline
        Enelvo~\cite{bertaglia2016exploring}  &  38.79\% & 15.84\%  & 19.39\% \\
        \hline
        Sridhar \cite{rangarajan-sridhar-2015-unsupervised}  & {\bf 39.79\%} & 16.20\%  & 19.91\% \\
        \hline
        Aspell  & 23.28\% & 9.68\%  & 12.44\% \\
        \hline
        \hline
        \model{} (Proposed)  & 39.23\% & {\bf 16.34\%}  & {\bf 20.01\%} \\
        %Significantly better than  &  - &  - & - & -\\
        \hline
    \end{tabular}
        \caption{{\bf Retrieval performance on TREC 2011 microblog dataset, using Indri IR system. Best performances are in bold font 
        %The proposed text normalization algorithm enables competitive retrieval performance, as compared to all baselines. 
        (none of the differences in performance are statistically significant).}}
    \label{tab:result-microblog}
    % \vspace{-5mm}    
\end{table}

\begin{table}[tb]
    \footnotesize
    \center
   % \addtolength{\tabcolsep}{-3pt}
    \begin{tabular}{|l|c|}
        \hline
        \textbf{Normalization Algorithm} & \textbf{Mean Reciprocal Rank} \\ 
        \hline
        \hline
        Raw data (without any cleaning)  & 62.93\%  \\
        \hline \hline
        Ghosh et al.\cite{ghosh-ipm} & 63.78\%  \\
        \hline
        Enelvo~\cite{bertaglia2016exploring} & 55.20\%  \\
        \hline
        Sridhar~\cite{rangarajan-sridhar-2015-unsupervised} & 54.36\% \\
        \hline
        Aspell & 38.93\%  \\
        \hline
        \hline
        \model{} (Proposed) & {\bf 70.21\%}$^{ESA}$  \\
       % Significantly better than &  A,E,S  \\
        \hline
    \end{tabular}
    \caption{{\bf Retrieval Performance on TREC 2005 Confusion dataset, using Indri IR system. Best result are in bold font. %The proposed \model{} method enables the best retrieval, compared to all baselines. Superscripts E,S,A indicate the proposed method is statistically significantly better than Enelvo, Sridhar and Aspell respectively.
    Superscripts E,S,A indicate the same as in Table~\ref{tab:result-risot}.
    }}    \label{tab:result-confusion}
    % \vspace{-10mm}    
\end{table}

% \begin{table}[tb]
%     \footnotesize
%     \center
%   % \addtolength{\tabcolsep}{-3pt}
%     \begin{tabular}{|l|c|}
%         \hline
%         \textbf{Normalization Algorithm} & \textbf{Mean Reciprocal Rank} \\ 
%         \hline
%         \hline
%         Raw data (without any cleaning)  & 32.68\%  \\
%         \hline \hline
%         Ghosh et al.\cite{ghosh-ipm} & 28.89\%  \\
%         \hline
%         Enelvo~\cite{bertaglia2016exploring} & \%  \\
%         \hline
%         Sridhar~\cite{rangarajan-sridhar-2015-unsupervised} & \% \\
%         \hline
%         Aspell & \%  \\
%         \hline
%         \hline
%         \model{} (Proposed) & {\bf \%}$^{ESA}$  \\
%       %  Significantly better than &  A,E,S  \\
%         \hline
%     \end{tabular}
%     \caption{{\bf Retrieval Performance on TREC 2005 Confusion (20\% noise) dataset, using Indri IR system. The best result is shown in bold font. The proposed \model{} method enables the best retrieval, compared to all baselines. Superscripts E,S,A indicate the proposed method is statistically significantly better than Enelvo, Sridhar and Aspell respectively.}}    \label{tab:result-confusion}
%     % \vspace{-10mm}    
% \end{table}

\subsection{Comparative Analysis}

Out of all the text normalization algorithms, retrieval using Aspell performs most poorly. 
%In fact, retrieval performance using Aspell normalization is worse than retrieval over the raw data, in most cases.
In general, spelling correction algorithms such as Aspell are unable to segregate words with high string similarity but low semantic similarity (`{\it industrious}' and `{\it industrial}'), or homonyms (such as `{\it living}' and `{\it leaving}'), etc.
Such variants can be identified by the proposed algorithm as well as by the other baselines which use context information in identification of variants.

In the rest of this section, we report some insights on how the proposed \model{} model differs from the other baselines Sridhar~\cite{rangarajan-sridhar-2015-unsupervised}, Enelvo~\cite{bertaglia2016exploring} and Ghosh et al.~\cite{ghosh-ipm}.
For this analysis, Table~\ref{tab:examples-microblog} and Table~\ref{tab:examples-confusion} show some examples of morphological variants identified by these algorithms for the same word.

%\vspace{2mm}
%\noindent \textbf{Proposed vs Aspell:} 
%The difference in the metrics of aspell and the proposed method shows that spelling correction algorithms like aspell cannot capture the morphological variants, e.g. homonymy. 
%In general, spelling correction algorithms such as Aspell are unable to segregate words with high string similarity but low semantic similarity ({\it industrious} and {\it industrial}), homonyms, etc.
%The low recall of Aspell is due to its lack of capability of query expansion. For example, for the TREC microblog dataset, the proposed model is able to capture words like {\it airport, airports} and put them in the same cluster, while there is no such way for Aspell. 
%Additionally, Aspell is unable to correct context sensitive errors such as {\it living} in ``{\it Please use the flush before living the toilet}''. Such errors can be identified by the other baselines and our proposed approach which uses context information in identification of variants.

\vspace{2mm}
\noindent \textbf{\model{} vs Sridhar/Enelvo:}
Both Sridhar and Enelvo {\it first} consider the contextual similarity for identifying a candidate set of morphological variants (see Eqn.~\ref{eqn:sridhar-step1} and Eqn.~\ref{eq:enelvo-knn} in the descriptions of Sridhar and Enelvo respectively). They initially ignore the structural/lexical similarity between words, which at times includes {\it incorrect} variants into the candidate set. 
For instance, the word `turmeric' is identified as a candidate variant of `organic' by Enelvo, while 
the words `nbcnews' and `tpmdc' are identified by both the methods as candidate variants of `msnbc' (see Table~\ref{tab:examples-microblog}), due to the high contextual similarity between these words. 
On the other hand, the proposed method considers $w_j$ to be a candidate variant of $w_i$ only if their lexical similarity is higher than a threshold; hence, such wrong variants are not included in $A(w_i)$ (on which contextual similarity is later applied).

\begin{table}[tb]
    \footnotesize
    \center
   % \addtolength{\tabcolsep}{-3pt}
    \begin{tabular}{|p{0.1\textwidth}|p{0.2\textwidth}|p{0.10\textwidth}|p{0.2\textwidth}|p{0.30\textwidth}|}
        \hline
        \textbf{Word} & \textbf{\model{}} &
        \textbf{Ghosh} &
        \textbf{Enelvo}& \textbf{Sridhar}\\ 
        \hline
        \hline
        release & release, released, releases & release & release, realease, relea & releasing, released, realease, release, releases, relea \\
        \hline
         organic & organic, organik, organi & organic & turmeric, organic & organic\\
         \hline
        % $Text_cleanser$ & -  \\
        % \hline
        msnbc & msnbc & msnbc & cbsnews, tpmdc, msnbc, foxnews, nbcnews & msnbc, nbcnews, cbsnews, tpmdc, masrawy \\
        \hline
        
    \end{tabular}
    \caption{{\bf Resulting clusters of morphological variants of some specific query-words, on the TREC 2011 microblog dataset}}
    \label{tab:examples-microblog}
    \vspace{-6mm}
\end{table}

\begin{table}[tb]
    \footnotesize
    \center
   % \addtolength{\tabcolsep}{-3pt}
    \begin{tabular}{|p{0.1\textwidth}|p{0.3\textwidth}|p{0.15\textwidth}|p{0.25\textwidth}|p{0.10\textwidth}|}
        \hline
        \textbf{Word} & \textbf{\model{}} &
        \textbf{Ghosh} &
        \textbf{Enelvo} & \textbf{Sridhar}\\ 
        \hline
        \hline
        % \hline
        program & programt, 44program, programs, program1, program programs1, sprogram, program11,  & program, programs & program programs1, program1, programs, prodects  & programs, program, programs1\\
        \hline
%         for & for  & for, form & for & for\\
%        \hline 
        % $Text_cleanser$ & -  \\
        % \hline
        document & document1, documents, documents1, locument1, locuments, locument, document, locuments1 & documents, document & document, forepart, locument, document1, documents & locument, documents, document1, document  \\
        \hline

    \end{tabular}
    
        \caption{{\bf Resulting clusters of morphological variants of some specific query-words, on the TREC 2005 Confusion dataset}}
    \label{tab:examples-confusion}
    \vspace{-10mm}    
\end{table}

\vspace{2mm}
\noindent \textbf{\model{} vs. Ghosh et al.:}
%Ghosh et. al~\cite{ghosh-ipm} uses $LCS$ as the similarity measure between two words to capture the syntactic similarity. 
Ghosh et al.~\cite{ghosh-ipm} computes the contextual similarity among words using only the co-occurrence counts of words in the same document. This method works well for long documents having a small vocabulary, but performs poorly when the vocabulary size is very large or when the document size is very small (in which case the co-occurrence matrix will be very sparse). 
Especially, two error variants of the same word rarely co-occur in the same microblog. 
%We argue that vocabulary size of noisy corpora are usually high thereby making co-occurrence counts inefficient to capture contextual information. For example the TREC 2011 microblog corpus contains total 2,821,480 unique tokens. 
Hence, Ghosh et. al~\cite{ghosh-ipm} performs well on the RISOT dataset, but poorly on the Microblog and Confusion datasets.
On the other hand, the proposed method employs both co-occurrence counts and cosine similarity of embeddings to capture contextual similarity among words, 
and hence generalizes better to different types of noise and documents.

\subsection{Ablation Analysis}

\begin{table}[tb]
    \footnotesize
    \center
    \addtolength{\tabcolsep}{2pt}
    \begin{tabular}{|l|c|c|c|}
        \hline
        \textbf{Normalization Algorithm} & \textbf{Recall@100} &\textbf{mAP@100} &\textbf{mAP@1000} \\ 
        \hline
        \hline
        \model{} (with all components)  & {\bf 58.3\%} & {\bf 21.5\%} & {\bf 23.0\%}\\
        \hline
        \hline
        \model{} without $\beta$ thresholding & 57.0 & 20.3 & 21.5 \\
        \hline  
        \model{} without similarity graph of morphological variants & 54.9 & 18.6 & 19.1 \\
        \hline        
        \model{} without graph fragmentation & 52.6 & 19.1 & 20.7 \\
        \hline
    \end{tabular}
        \caption{{\bf Ablation analysis of \model{} -- how retrieval performance (on RISOT Bengali OCR dataset, using Indri IR system) varies with removal of various components.}}
    \label{tab:ablation-risot}
    % \vspace{-10mm}    
\end{table}

We conducted an ablation analysis in order to understand the importance of various steps / components in our algorithm. To this end, we applied \model{} on the RISOT dataset in various settings -- with all steps, and by removing one step at a time.
Table~\ref{tab:ablation-risot} shows the results of the ablation analysis.

The $\beta$ thresholding (see Step 2 in Section~\ref{sub:algo-construct-sim-graph}) is done to consider words which have a high contextual similarity only. This is done to ensure high precision of the retrieved results. As can be seen from Table~\ref{tab:ablation-risot}, the removal of such a threshold results in the drop of mAP scores.
The recall@100, however, does not reduce much, since the morpheme cluster increases.

The construction of similarity graph of morphological variants (see Step 2 in Section~\ref{sub:algo-construct-sim-graph}) is a crucial step that helps capture the contextual similarity among words with different morphological structures. Removal of this step will result in a significant drop in both  mAP and recall scores compared to \model{}.

If the graph fragmenting step (see Step 3 in Section~\ref{sub:algo-fragment-graph}) is skipped, we miss out on identifying the strongly connected component of the graph, and thus weakly connected components also get drawn into the final cluster, causing a decrease in both mAP as well as recall scores.

This ablation analysis shows that all the steps of the \model{} algorithm are essential, and removing any of the steps will adversely affect the retrieval performance. 
Especially, skipping the graph formation step leads to substantial degradation in both Precision and Recall of retrieval.
Also the graph fragmenting step is crucial particularly for Recall.

\vspace{2mm}
\noindent From the experiments and analyses stated in this section, we can conclude the our proposed algorithm \model{} is very helpful in enhancing retrieval from noisy text.
\model{} out-performs almost all baselines across different types of noisy corpus (containing OCR noise, and noise due to cavalier writing style of human users) and across multiple languages (English and Bengali).
Especially for corpus with OCR noise, \model{} enables statistically significantly better retrieval than most baselines, while it achieves very competitive performance in case of noisy microblogs.
Additionally, \model{} is completely unsupervised, and does not need any manual intervention at any step, thus making it suitable for use on large corpora in various languages.

%\textcolor{blue}{ From the above experiments and analysis it can be seen that information retrieval approaches depend heavily on the morphological variations of query words. For example in table ~\ref{tab:examples-microblog} for the query word `organic`, documents containing its morphological variants like `organik, organi` should also be retrieved. Such morphological variations of the query word are nicely captured by the \model{}. Words like `turmeric` captured by Enelvo, however, is not a morphological variant of organic. This happens because of the cosine similarity based KNearest Neighbors~\ref{eq:enelvo-knn} as the initial method for finding similar variants. The score computation in the next step also played a role in this wrongful identification of morphological variants. In The scoring function of Enelvo~\ref{eq:enelvo-score}, the high contextual similarity(i.e. cosine similarity) between turmeric and organic outweighed the low lexical similarity between them. Even if contextual similarity part is dropped in the scoring function like in Sridhar et al., the high lexical similarity among those that are contextually similar may lead to wrong results like in the case of the query word `msnbc`.}

\section{Downstream Application 2: Stance Detection} \label{sec:stance-detect}

%\textcolor{blue}{ TODO -- Why is stance detection a good downstream application for evaluation of the models? Stance detection is a task for which there exist some standard techniques. So we wanted to see if we apply our preprocessing technique and then detect stance using the standard methods whether the proposed normalization method performs better compared to when other preprocessing methods are applied. In that way we will be able to check how effective the proposed algorithm is for a task like stance detection if applied on multiple models(Here TAN and LSTM). It might be the case that for a particular stance detection method the proposed normalization method outperforms other existing normalization techniques and does not work better for a different stance detection algorithm (indicating the strength of the stance detection algorithm). So we have chosen two stance detection methods and will try to see whether across both the methods and majority of the datasets the proposed model outperforms other models,thus the effectiveness of the proposed method will be proved.}

Online platforms such as Twitter, Facebook, etc. have become popular forums to discuss about various socio-political topics and express personal opinions regarding these topics. 
In this kind of scenario, `stance' refers to the inherent sentiment express by an opinion towards a particular topic. The topic is referred to as the `target'. The `stance' expressed by a blog/tweet can have three types of sentiment towards a given topic. It can either {\it support} (or {\it favor}) the target, it can {\it oppose} (or be {\it against}) the target, or it can be {\it neutral}, i.e., neither support nor oppose the target. The problem of `stance detection' refers to the process of automatically identifying this stance or sentiment of a post expressed towards a target. The reader is referred to~\cite{stance-detection-survey} for a recent survey on stance detection. 

%Prior works show that stance detection can be of two types,(i)~Multi-target stance detection and (ii)~Target specific stance detection. In this work the models chosen are applied for single target stance detection.

Stance detection is frequently carried out on crowdsourced data such as tweets/blogs which are noisy in nature, due to frequent use of informal language, non-standard abbreviations, and so on.
There exist several popular methods for stance detection from such noisy crowdsourced content.
%\textcolor{blue}{For this task there exist some standard techniques. So we wanted to see if we apply our preprocessing technique and then detect stance using the standard methods whether the proposed normalization method performs better compared to when other preprocessing methods are applied. In that way we will be able to check how effective the proposed algorithm is for a task like stance detection if applied on multiple models(Here TAN and LSTM). It might be the case that for a particular stance detection method the proposed normalization method outperforms other existing normalization techniques and does not work better for a different stance detection algorithm (indicating the strength of the stance detection algorithm). So we have chosen two stance detection methods and will try to see whether across both the methods and majority of the datasets the proposed model outperforms other models,thus the effectiveness of the proposed method will be proved.}
In this section, we check whether using text normalization algorithms improves the performance of these stance detection models on noisy microblogs.

\subsection{Datasets for stance detection}

We consider datasets made available by the SEMEVAL 2016 Task 6A challenge~\cite{mohammad2016semeval} which are commonly used datasets for stance detection, and have been used in a number of prior works. The datasets consist of tweets (microblogs) for different topics; for each topic, there are three types of tweets -- those in favor of (supporting) the topic, those against (opposing) the topic, and tweets that are neutral to the topic.
We consider the following three SemEval datasets:\\
(i)~\textbf{Atheism (AT)}:  Here the target is `atheism', and the tweets are in favor of, or against the idea of atheism (or neutral). \\
(ii)~\textbf{Climate change is a real concern (CC)}: For this dataset the target is `Climate change is a real concern'.\\
(iii)~\textbf{Hillary Clinton (HC)}: Here the target is `Hillary Clinton', and the tweets are either in support of / against the politician.\\
We refer to these datasets as AT, CC and HC respectively for brevity in the rest of the paper. 

%For all the above 3 datasets there can be 3 types of tweets based on type of opinion expressed:
%\begin{enumerate}
%    \item   The tweet explicitly expresses opinion about the target, a part of the target, or an aspect of the target.
%    \item  The tweet does NOT expresses opinion about the target but it HAS opinion about something or someone other than the target.
%    \item   The tweet is not explicitly expressing opinion. (For example, the tweet can simply giving information.)
%\end{enumerate}

The datasets are already partitioned into train and test sets, as defined in the SEMEVAL 2016 Task 6A challenge~\cite{mohammad2016semeval}. 
The numbers of tweets in each dataset is stated in Table~\ref{tab:dataset-stance}. 
We have also provided some examples of tweets in the datasets in Table~\ref{tab:examples-stance}. 

\begin{table}[tb]
\center
\footnotesize
%\resizebox{!}{35pt}{
\begin{tabular}{|c||c|c|c||c|c|c|}
\hline
\textbf{Dataset} & \multicolumn{3}{c||}{\textbf{\# Training Instances}}  & \multicolumn{3}{c|}{\textbf{\# Test Instances}}\\
\hline
 &  Favor & Against & Neutral &  Favor & Against & Neutral \\
\hline
SemEval-AT & 92 & 304 & 117 & 32 & 160 & 28 \\ \hline
SemEval-CC & 212 & 15 & 168 & 123 & 11 & 35 \\ \hline
%SemEval-FM & 210 & 328 & 126 & 58 & 183 & 44 \\ \hline
SemEval-HC & 112 & 361 & 166 & 45 & 172 & 78 \\ 
\hline
\end{tabular}
%}
\caption{\textbf{Statistics of standard SEMEVAL 2016 Task 6A datasets for stance detection (divided into training and test sets).}}
\label{tab:dataset-stance}
\end{table}

\begin{table*}[tb]
\footnotesize
\centering
    \begin{tabular}{|p{0.85\textwidth}|c|} \hline
    \textbf{Tweet text} & \textbf{Label}\\ \hline
    \multicolumn{2}{|c|}{{\bf Tweets from SemEval-AT (Atheism) dataset}} \\ \hline

   Absolutely f**king sick \& tired of the religious and their "We're persecuted" bollocks\! So f**king what? Pissoff! \#SemST & FAVOR \\ \hline
        
    All that is needed for God for something to happen is to say "\#Be" and it is; for God is capable of all things. \#God created \#trinity \#SemST & AGAINST\\ \hline
        
    In other related news. Boko Haram has killed over 200 people in the last 48hrs. \#SemST & NEUTRAL  \\ \hline
    
     \multicolumn{2}{|c|}{{\bf Tweets from SemEval-CC (Climate Change is a Real Concern) dataset}} \\ \hline
    
    We cant deny it, its really happening.  \#SemST & FAVOR\\ \hline
    
    The Climate Change people are disgusting a**holes. Money transfer scheme for elite. May you rot.   \#SemST & AGAINST\\ \hline
    
    @AlharbiF I'll bomb anything I can get my hands on, especially if THEY aren't christian. \#graham2016 \#GOP \#SemST & NEUTRAL\\ \hline
    
    \multicolumn{2}{|c|}{{\bf Tweets from SemEval-HC (Hillary Clinton) dataset}} \\ \hline
    
    @HuffPostPol If @HillaryClinton can do half of what he did then she would be doing is a favor \#SemST & FAVOR\\ \hline

    Although I certainly have disagreements, after reading about @GovernorOMalley I much rather have him than @HillaryClinton. \#SemST & AGAINST\\ \hline

    Lets remember \#dickcheney is an unindicted war criminal b4 we start yelling \#Benghazi day after day. Will we ever see justice? \#SemST & NEUTRAL\\ \hline

    \end{tabular}
    \caption{\textbf{Examples of posts from the SemEval datasets. Some characters in abusive words are replaced with *}}
    \label{tab:examples-stance}
    \vspace{-5mm}
\end{table*}

\subsection{\bf Stance Detection Models}

We consider two stance detection models for the experiments. 

\begin{enumerate}

\item \textbf{LSTM (Long Short Term Memory)}: This model was proposed by Hochreiter and Schmidhuber~\cite{hochreiter1997long} to specifically address this issue of learning long-term dependencies. 
The LSTM maintains a separate memory cell inside it that updates and exposes its content only when deemed necessary.A number of minor modifications to the standard LSTM unit have been made. 
We define the LSTM units at each time step t to be a collection of vectors in
$R^{d}$ : 
an input gate $i_{t}$,
a forget gate $f_{t}$, 
an output gate $o_{t}$,
a memory cell $c_{t}$,
and a hidden state $h_{t}$ .
$d$ is the number of the LSTM units. The entries of the gating vectors
$i_{t}$, $f_{t}$ and $o_{t}$ are in $[0,1]$. The LSTM transition equations are the following: 
\begin{equation}
i_{t} = \sigma (W_{i}x_{t}+U_{i}h_{t-1}+V_{i}c_{t-1})    
\end{equation}
\begin{equation}
f_{t} = \sigma (W_{f}x_{t}+U_{f}h_{t-1}+V_{f}c_{t-1})    
\end{equation}
\begin{equation}
o_{t} = \sigma (W_{o}x_{t}+U_{o}h_{t-1}+V_{o}c_{t-1})    
\end{equation}
\begin{equation}
\tilde{c}_{t} = \sigma (W_{i}x_{t}+U_{i}h_{t-1}+V_{i}c_{t-1})    
\end{equation}
\begin{equation}
c_{t} = f_{t} \odot c_{t-1} + i_{t} \odot \tilde{c}_{t}     
\end{equation}
\begin{equation}
h_{t} = o_{t} \odot tanh(c_{t})      
\end{equation}

For detecting stance , a simple strategy is to map the input sequence to a fixed-sized vector using one RNN, and then to feed the vector to a softmax layer. Given a text sequence, x=$[x_{1},x_{2}$,···$,x_{T}]$,  we first use a
lookup layer to get the vector representation (embeddings)
of each word $\mathbf{x}_{i}$. The output $\mathbf{h}_{T}$ can be regarded as the representation of the whole sequence, which has a fully connected layer followed by a softmax non-linear layer that predicts the probability distribution over stances (in this case either `FAVOR' or `AGAINST' or `NEUTRAL').
The LSTM model has the following hyperparameters whose values are taken as follows: learning rate of $5 \times e^{-4}$ and dropout of $0.5$.

\vspace{5mm}

\item \textbf {Target Specific Attention Neural Network (TAN)}: This model was introduced by Du et al.~\cite{Du:2017:SCT:3171837.3171843}, one of the winning entries SemEval 2016 task 6A~\cite{mohammad2016semeval} challenge. The model is based on bidirectional LSTM with an attention mechanism. A target sequence of length $N$ is represented as $[z_1, z_2, \ldots , z_N]$
where $z_n \epsilon R^{d^{'}}$ is the $d^{'}$-dimensional vector of the $n$-th word in the target sequence. 
The target-augmented embedding of a word $t$
for a specific target $z$ is $e^{z}_{t}=x_{t} \odot z$ where $\odot$ is the vector concatenation operation. 
The dimension of $e^z_t$ is $(d+d^{'})$. An affine transformation maps the $(d+d^{'}$)-dimensional target-augmented embedding of each word to a scalar value as per the following Eqn.~\ref{eqn:eqn13}: 
\begin{equation}
a^{'}_{t} = W_{a}e_{t}^{z} + b_{a}   
\label{eqn:eqn13}
\end{equation} 
where $W_a$ and $b_a$ are the parameters of the bypass neural network. The attention vector $[a^{'}_{1}, a^{'}_{2}, \ldots, a^{'}_{T}]$ undergoes a softmax transformation to get the final attention signal vector (Eqn.~\ref{eqn:eqn14}):
\begin{equation}
    a_{t} = softmax(a_{t}) = \frac{e^{a^{'}_{t}}}{\sum_{i=1}^{T}e^{a^{'}_{i}}}
    \label{eqn:eqn14}
\end{equation}

After this, the product of attention signal
$a_t$ and $h_t$ (which is the corresponding hidden state vector of RNN) is used to represent the word t in a sequence with attention signal.  The representation of the whole sequence can be obtained by averaging the word representations:

\begin{equation}
    s = \frac{1}{T}\sum_{t=0}^{T}a_th_t
\end{equation}
 
where $s$ $\epsilon$ $R^d$ is the vector representation of the text sequence and it can be used as features for text classification: 

\begin{equation}
    p = softmax(W_{clf}s + b_{clf})
\end{equation}

where
$p$ $\epsilon$ $R^C$ is the vector of predicted probability for stance. Here $C$
is the number of classes of stance labels, and
$W_{clf}$ and $b_{clf}$ are parameters of the classification layer. 

The TAN model has two hyperparameters, whose values are taken as follows: learning rate of $5 \times e^{-4}$ and dropout of $0.5$.

%Figure ~\ref{TAN} shows the basic architecture of the target specific attention neural network~\cite{Du:2017:SCT:3171837.3171843}.

%\begin{figure}[H]
%\centering
%\includegraphics[width = \linewidth]{Capture_TAN.PNG}.
%\caption{Main architecture of the TAN model ~\cite{Du:2017:SCT:3171837.3171843}}
%\label{TAN}
%\end{figure}

\end{enumerate}

\subsection{Parameter settings of normalization algorithms} \label{sub:stance-parameter-setting}

We set the parameters for various text normalization algorithms in a way similar to what was described in Section~\ref{sub:retrieval-parameter-setting}. 
For the proposed method, the value of $\alpha$ is set of $0.56$. 
For Ghosh et al.~\cite{ghosh-ipm}, the parameters were set to $\alpha = 0.7, \beta = 0.6, \gamma = 50$ (decided through grid search, as described in Section~\ref{sub:retrieval-parameter-setting}).
For Sridhar~\cite{rangarajan-sridhar-2015-unsupervised} and Enelvo~\cite{bertaglia2016exploring}, the hyperparameter $K$ was set to $25$ as specified in the original papers.
For Enelvo~\cite{bertaglia2016exploring}, $n$ was set to $0.8$.

\subsection{Evaluation Metric for stance detection} \label{sub:stance-metric}

To evaluate the performance of different models, we use the same metric as reported by the official SemEval 2016 Task A~\cite{mohammad2016semeval}. 
We use the macro-average of the F1-score for `favor' and `against' as the bottom-line evaluation metric.
    \begin{equation}
    F_{avg} =\frac{F_{favor} + F_{against}}{2}   
    \end{equation}
    where $F_{favor}$ and $F_{against}$ are calculated as shown below:
    \begin{equation}
    F_{favor} = \frac{2P_{favor}R_{favor}}{P_{favor}+R_{favor}}
    \end{equation}
   
   \begin{equation}
    F_{against} =\frac{2P_{against}R_{against}}{P_{against}+R_{against}}
    \end{equation}
Here $P_{favor}$ and $R_{favor}$ are the precision and recall of the `FAVOR' class respectively,  and $P_{against}$ and $R_{against}$ respectively are the precision and recall of the `AGAINST' class and are defined as:
    \begin{equation}
        P_{favor} = \frac{\#TP_{favor}}{\#TP_{favor}+\#FP_{favor}}
    \end{equation}
    
    \begin{equation}
        R_{favor} = \frac{\#TP_{favor}}{\#TP_{favor}+\#FN_{favor}}
    \end{equation}
    
    \begin{equation}
        P_{against} = \frac{\#TP_{against}}{\#TP_{against}+\#FP_{against}}
    \end{equation}
    
    \begin{equation}
        R_{against} = \frac{\#TP_{against}}{\#TP_{against}+\#FN_{against}}
    \end{equation}
With respect to `FAVOR' class, \(\#TP_{favor}\) refers to number of true positives in favor class, i.e., the number of tweets that are predicted as `FAVOR' by the classifier, and are actually of `FAVOR' class. 
\(\#FP_{favor}\) refers to number of tweets which are not actually of `FAVOR' class, but that have been predicted as `FAVOR' by the classifier. 
\(\#FN_{favor}\) refers to number of tweets which are actually of `FAVOR' class but have been wrongly classified by the classifier.

\subsection{Results}

The experiments with a particular dataset are as follows. Five instances of the dataset are normalized, one by each of the five normalization algorithms (\model{} and four baselines).
The training set and the test set are normalized separately.
Subsequently, a stance detection model (LSTM or TAN) is used on the normalized versions, and their performances are measured. 

Table~\ref{table:twitter_result_TAN} shows the performance of the TAN stance detection model~\cite{Du:2017:SCT:3171837.3171843}, on different versions of the datasets normalized by the different normalization algorithms.
Also shown are the values on the raw datasets, i.e., without any cleaning.
From Table~\ref{table:twitter_result_TAN}, it can be seen that all normalization approaches lead to some improvement, as compared to the performance on the raw data.
It is evident that the TAN model performs the best by using our proposed normalization method, for all three datasets (best metric values highlighted in boldface). 

Similarly, Table~\ref{table:twitter_result_LSTM} shows the performance of the LSTM~\cite{hochreiter1997long} model
on different versions of the datasets normalized by the different normalization algorithms, as well as on the raw dataset.
The LSTM model performs the best by using the proposed normalization method for two datasets (AT and CC), and gives the second-best result for the HC dataset. 

%It can be noted that Aspell performs competitively with the other normalization algorithms over these stance detection datasets, unlike in the case of retrieval where Aspell performed very poorly. This difference in performance is possibly because SemEval datasets have less noise, and the noisy variants are closer to the original words. 

We checked the statistical significance of the differences in performance with various normalization methods using McNemar's test (\url{https://machinelearningmastery.com/mcnemars-test-for-machine-learning/}); 
however, the differences in performance are not statistically significant.

%Next, to check the statistical significance of the model we performed McNemar's test (\url{https://machinelearningmastery.com/mcnemars-test-for-machine-learning/}) for comparing classifiers based on the predicted labels by different baselines as well as the proposed method. We found that the proposed method is not statistically significant from baseline methods w.r.t P value. Also from the literature of `Stance detection' we can say that none of the well known methods established so far goes for statistical significance testing. One reason might be the metric used for evaluation of stance detection models as mentioned by SemEval challenge is not adaptable to tests like McNemar's test or paired student's t test. It might require some changes in the test setup.  

%\kripa{Significance tests are possible for classification also, e.g. see \url{cmpe.boun.edu.tr/~ethem/i2ml/slides/v1-1/i2ml-chap14-v1-1.pdf}. In the paper \url{http://www.bigdatalab.ac.cn/~gjf/papers/2019/SIGIR-crime.pdf} (see Table 5), t-test results have been reported. I have never done it myself and hence don't have the code. It will look odd if t-tests are reported for retrieval and not for stance classification.}

\begin{table*}[tb]
\centering
\scalebox{0.85}{
\begin{tabular}{|p{0.4\textwidth}||c|c|c|c|}
\hline
 \textbf{Normalization Method}&\textbf{AT} & \textbf{CC} &  \textbf{HC} \\ \hline
Raw data (without any cleaning) & 0.5508 & 0.3968  & 0.4396 \\ \hline \hline
Enelvo~\cite{bertaglia2016exploring} &  0.5659 & 0.4427 & 0.5336  \\ \hline
Sridhar~\cite{rangarajan-sridhar-2015-unsupervised} &  0.5476 & 0.4778 & 0.5318 \\ \hline

Ghosh et al.~\cite{ghosh-ipm}. &  0.6156 & 0.4641 & 0.5216 \\ \hline

Aspell &  0.5923 & 0.5250 & 0.4586\\ \hline\hline
\model{} (Proposed)  & {\bf 0.6250} & {\bf 0.5271} & {\bf 0.5405}\\ \hline
\end{tabular}}
\caption{{\bf Stance detection results on SemEval datasets by the TAN model~\cite{Du:2017:SCT:3171837.3171843}. The metric is as explained in Section~\ref{sub:stance-metric}. Highest values marked in boldface.}}
\label{table:twitter_result_TAN}
%\vspace*{-5mm}
\end{table*}

\begin{table*}[tb]
\centering
\scalebox{0.85}{
\begin{tabular}{|p{0.4\textwidth}||c|c|c|c|}
\hline
 \textbf{Normalization Method}&\textbf{AT} & \textbf{CC} &  \textbf{HC} \\ \hline
Raw data (without any cleaning) &  0.5497 & 0.3760 & 0.4938  \\ \hline \hline
Enelvo~\cite{bertaglia2016exploring} &  0.6084 & 0.3911 & {\bf 0.5646} \\ \hline
Sridhar~\cite{rangarajan-sridhar-2015-unsupervised} &  0.5666 & 0.4558 & 0.5055 \\ \hline

Ghosh et al.~\cite{ghosh-ipm}. &  0.6395 & 0.3884 & 0.4893 \\ \hline

Aspell &  0.5902 & 0.4008 & 0.5337 \\ \hline\hline
\model{} (Proposed)  & {\bf 0.6411} & {\bf 0.4882} & 0.5401  \\ \hline
\end{tabular}}
\caption{{\bf Stance classification results on SemEval datasets by the LSTM~\cite{hochreiter1997long} model.  The metric is as explained in Section~\ref{sub:stance-metric}. Highest values marked in boldface.}}
\label{table:twitter_result_LSTM}
%\vspace*{-5mm}
\end{table*}

\subsection{Analysis of result of stance detection}

%\textcolor{blue}{While detecting stance we need to preprocess a tweet or a text. A sentence might contain some out of vocabulary words which need to addressed and also some representation of those words are required while applying some algorithm on that sentence. Now there are many supervised methods to address OOV words. The main problem with those methods are they might not be able to capture the context which is very important for detecting stance. The proposed method will try to normalize a OOV word to its closest word based on the training corpus.}

Table~\ref{tab:fail} shows some examples of tweets from the SemEval datasets, where all/most of the baseline normalization algorithms led to {\it wrong} stance detection by the TAN~\cite{Du:2017:SCT:3171837.3171843} model (for the first two examples) and by the LSTM  model~\cite{hochreiter1997long} (for the last two examples), but by using the proposed normalization method, the correct stance was identified for each of these tweets.
%The first two examples are from the results produced by the TAN~\cite{Du:2017:SCT:3171837.3171843} stance detection model and last two examples are from the results produced by the LSTM model~\cite{hochreiter1997long}.

The first and third  examples show a situation where  the proposed \model{} keeps the words present in a tweet unchanged, but the baseline normalization methods change the words to some non-contextual variants that seem illogical and lead to wrong stance detection. 
In the second example the proposed method changes the phrase `thank you so much for' to `thank you so much your' whereas the modifications of different words by the other normalization methods seem wrong, such as from `@EstadodeSats' to `stardust's' by Aspell, or from `thank' to 'thang' by Enelvo. 
A similar scenario is observed in the fourth example.
Thus, the proposed model performs slightly better in understanding which words to modify (into variants) and which words to leave unchanged. 
%Specially, the aspell model it changes all the `#' symbols to `W' which leds to wrong embedding resulting in wrong prediction by the aspell model.

\vspace{2mm}
\noindent From the experiments described in this section, it can be concluded that the proposed text normalization algorithm \model{} competes very favorably with the baseline normalization models, for the stance detection task. 
We demonstrate the efficacy of \model{} with two popular stance detection models. Experiments show that both stance detection models perform better after text normalization with \model{}, as compared to text normalization with the baseline models.

These results, along with the results of retrieval in the previous Section~\ref{sec:expt-retrieval}, show that text normalization using \model{} enables superior performance in multiple text processing tasks.

\begin{table*}[tb]
\footnotesize
\centering
\begin{tabular}{|p{0.30\textwidth}|p{0.10\textwidth}|p{0.10\textwidth}|p{0.10\textwidth}|p{0.10\textwidth}|p{0.10\textwidth}|} \hline
        \textbf{Tweets from SemEval Datasets} & \textbf{UnsupClean (Proposed)} & \textbf{Enelvo~\cite{bertaglia2016exploring}} & \textbf{Sridhar~\cite{rangarajan-sridhar-2015-unsupervised}}  & \textbf{Ghosh et al.~\cite{ghosh-ipm}}& \textbf{Aspell} \\ \hline
    %\multicolumn{2}{|c|}{{\bf Tweets from SemEval Dataset AT (Atheism)}} \\ \hline
       {\bf Jesus} response to a {\bf religious} environment  was to {\bf create} a royal environment.  It drove people nuts. Still does - thank God \#SemST
 & jesus, \newline religious, \newline create & janus, \newline Seditious, \newline create & joss, \newline regs, \newline create & jesus, \newline religious, \newline creative & jesus, \newline religious, \newline create, \newline replaces `\#' by `W'\\ \hline

       %@HillaryClinton : "Change religious beliefs" to accommodate the violence of abortion?! \#WhyI'mNotVotingForHillary \#SemST

 %& hillaryclinton & next & illegality &   & hydrogenating, replaces `\#' by `W'\\ \hline
       
    @csham21 {\bf @EstadodeSats} {\bf Thank you} so {\bf much} {\bf for} supportive \#Freedom4Cuba RTs! \#WakeUpAmerica \#Cuba \#TodosMarchamos \#SemST

 & estadodesats, \newline  thank you,  \newline much your & estadodesats,  \newline   thang Lou,  \newline  guch onr & estadodesats,  \newline  thanks yU,  \newline mce onr & estadodesats,  \newline  anything 4your, \newline  much 4your & stardust's, \newline  thank you, \newline  much for, \newline replaces `\#' by `W'\\ \hline

    .@mite72 {\bf @PatVPeters} NEITHER {\bf ONE}!!!! {\bf Just making} a {\bf funny!} \#SemST & 
     patvpeters, \newline one, \newline just making funny  & patvpeters, \newline obe, \newline sunt making  Hanno & 
     patvpeters, \newline ne, \newline Fst making  fane &  patvpeters, \newline doesn,\newline just making  tofunny  & parapets \newline one, \newline just  making  funny, \newline replaces `\#' and `\!' by `W'\\ \hline

    {\bf\#Religions} can't {\bf all be right, but they can all be wrong.} \#SemST & \#religions,\newline  all be right,\newline but their can all be wrong & \#religious \newline alt be Light, Wut theo caf alt be Frond & \#riggs, \newline Fall be rights, bte thay Un Fall be Lrg & \#religions \newline all be right, but nothing can all be wrong & 	  religions \newline all Be Right but they Can all Be wrong, \newline replaces `\#' by `W' \\ \hline

    %@Brasilmagic @neiltyson there's no god 2b tired of, just tired o ppl believing in such nonsense. Time 2 focus on reality, HUMANITY \#SemST  & & & &  &\\ \hline

    \end{tabular}
    \caption{\textbf{Examples of posts from the SemEval datasets, for which proposed method led to correct stance detection but all/most baseline normalization methods led to wrong stance detection (by TAN~\cite{Du:2017:SCT:3171837.3171843} for first two examples, and by LSTM model~\cite{hochreiter1997long} for last two examples). 
    %Shown are the original tweets and variants of words after normalization by different algorithms. The first two examples are from the results produced by TAN~\cite{Du:2017:SCT:3171837.3171843} stance detection model and last two examples are from the results produced by the LSTM model~\cite{hochreiter1997long}.
    }}
    \label{tab:fail}
\end{table*}

% \vspace{-3mm}
\section{Conclusion}\label{sec:conclusion}
% \vspace{-3mm}

We proposed a language-independent, unsupervised algorithm for normalizing/cleansing of noisy text. 
%Experimental results show that the proposed method generalizes better than several baselines to different types of noise (machine-generated OCR noise and user-generated noise in microblogs). 
%While existing methods perform well for one type of noise (e.g., Sridhar and Enelvo for noise in microblogs, and Ghosh et al. for OCR noise),
We conducted experiments for two downstream applications (Retrieval and Stance detection), over a variety of datasets and types of noise, including OCR noise over English and Bengali, as well as noise due to informal writing style of humans in microblogs. 
The experiments show that the proposed method generalizes well to different types of noise (both user-generated and machine-generated noise), and performs competitively with several baseline text normalization algorithms. 
The main strengths of the proposed method are that (i)~it does not need expensive parallel corpus for training or any human intervention (unlike many existing algorithms), and
(ii)~it does not need external resources such as global word embeddings. These features make it any attractive choice for cleaning text in low-resource languages. 
The implementation of \model{} is publicly available at \url{ https://github.com/ranarag/UnsupClean}.

There are several potential future directions of this work. First, the effectiveness of \model{} can be checked on other types of noise, such as noise due to automated speech-to-text conversion systems. 
Second, the proposed method can be applied to clean text in low-resource languages that lack external resources and large parallel corpora. 
Also, in this paper, we demonstrated the benefits of cleaning text using \model{} for the two general tasks of retrieval (search) and stance detection. 
The proposed method can also be tried to improve performance in more specific versions of these tasks, such as identifying specific types of microblogs that aid relief operations during post-disaster operations~\cite{tcss_Basu}, 
as well as in other tasks such as summarization of noisy social media text~\cite{rudra-cikm-disaster,social-media-opinion-summarize}. 
We plan to explore these directions in future.

\vspace{3mm}
\noindent {\bf Acknowledgements:} The authors thank the anonymous reviewers whose valuable suggestions helped to improve the paper. 
The authors also acknowledge Dr. Arnab Bhattacharya of Indian Institute of Technology Kanpur for useful discussions in the initial stages of the work.
The work is partially supported by a project titled ``Building
Healthcare Informatics Systems Utilising Web Data'' funded by Department of Science
\& Technology, Government of India.
Finally, the authors gratefully acknowledge the support of NVIDIA Corporation with the donation of the Titan Xp GPU used for this research.

\bibliographystyle{unsrt}

\end{document}